\documentclass[10pt,twocolumn,twoside,journal]{IEEEtran}

\newcommand{\Define}{\stackrel{\triangle}{=}}
\newcommand{\argmin}{\operatornamewithlimits{argmin}}

\usepackage{cite}
\usepackage{latexsym}
\usepackage{epsfig}
\usepackage{verbatim}
\usepackage{amssymb}
\newtheorem{thm}{\bf Theorem}

\usepackage{graphicx}
\usepackage{color}
\usepackage[cmex10]{amsmath}

\usepackage{subfigure}
\usepackage{algorithmic}
\usepackage{algorithm}
\usepackage{multirow}

\begin{document}

\title{{\huge Generalized Space and Frequency Index Modulation}}
\author{T. Datta, H. S. Eshwaraiah, and A. Chockalingam, 
\thanks{Tanumay Datta is presently with Centrale Supelec, Gif sur Yvette,
France 91190. E-mail: tanumaydatta@gmail.com. 

Harsha S. Eshwaraiah did this work when he was with the Department
of Electrical Communication Engineering, Indian Institute of Science, 
Bangalore, India. E-mail: seharsha@gmail.com. 

A. Chockalingam is with the Department of Electrical Communication 
Engineering, Indian Institute of Science, Bangalore-560012, India.
E-mail: achockal@ece.iisc.ernet.in.}
}
\date{}

\maketitle

\begin{abstract}
Unlike in conventional modulation where information bits are conveyed
only through symbols from modulation alphabets defined in the complex
plane (e.g., quadrature amplitude modulation (QAM), phase
shift keying (PSK)), in index modulation (IM), additional information
bits are conveyed through indices of certain transmit entities that
get involved in the transmission. Transmit antennas in multi-antenna
systems and subcarriers in multi-carrier systems are examples of such
transmit entities that can be used to convey additional information
bits through indexing. In this paper, we introduce {\em generalized
space and frequency index modulation}, where the indices of active
transmit antennas and subcarriers convey information bits. We first
introduce index modulation in the spatial domain, referred to as
generalized spatial index modulation (GSIM). For GSIM, where bits are
indexed only in the spatial domain, we derive the expression for
achievable rate as well as easy-to-compute upper and lower bounds on
this rate. We show that the achievable rate in GSIM can be more than
that in spatial multiplexing, and analytically establish the condition
under which this can happen. It is noted that GSIM achieves this higher
rate using fewer transmit radio frequency (RF) chains compared to spatial
multiplexing. We also propose a Gibbs sampling based detection algorithm
for GSIM and show that GSIM can achieve better bit error rate (BER)
performance than spatial multiplexing. For generalized space-frequency
index modulation (GSFIM), where bits are encoded through indexing in
both active antennas as well as subcarriers, we derive the achievable
rate expression. Numerical results show that GSFIM can achieve higher
rates compared to conventional MIMO-OFDM. Also, BER results show the
potential for GSFIM performing better than MIMO-OFDM.
\end{abstract}

\begin{keywords}
Multi-antenna systems, multi-carrier systems, spatial index modulation,
space-frequency index modulation, achievable rate, transmit RF chains,
detection.
\end{keywords}

\section{Introduction}
\label{sec1}
Multi-antenna wireless systems have become very popular due to their high
spectral efficiencies and improved performance compared to single-antenna
systems \cite{tse}-\cite{lmimo}. Practical multi-antenna systems are faced
with the problem of maintaining multiple radio frequency (RF) chains at the
transmitter and receiver, and the associated RF hardware complexity, size,
and cost \cite{rf}. Spatial modulation, a transmission scheme which uses
multiple transmit antennas but only one transmit RF chain, can alleviate
the need for multiple transmit RF chains \cite{SM2}-\cite{lajos}.
In spatial modulation, at any given time, only one among the transmit
antennas will be active and the other antennas remain silent. The index
of the active transmit antenna will also convey information bits, in
addition to the information bits conveyed through the conventional
modulation symbol (e.g., chosen from QAM/PSK alphabet) sent on the active
antenna. An advantage of spatial modulation over conventional modulation
is that, for a given spectral efficiency, conventional modulation requires
a larger modulation alphabet size than spatial modulation, and this can
lead to spatial modulation performing better than conventional modulation
\cite{mu1},\cite{mu3}.

In this paper, we take the view that spatial modulation is an instance of
the general idea of `index modulation'. Unlike in conventional modulation
where information bits are conveyed only through symbols from modulation
alphabets defined in the complex plane (e.g., QAM, PSK), in index modulation
(IM), additional information bits are conveyed through indices of certain
transmit entities that get involved in the transmission. Transmit antennas
in multi-antenna systems, subcarriers in multi-carrier systems, and
precoders are examples of such transmit entities that can be used to
convey information bits through indexing. Indexing in spatial domain
(e.g., spatial modulation, and space shift keying which is a special case
of spatial modulation) is a widely studied and reported index modulation
technique; see \cite{lajos} and the references therein. Much fewer works
have been reported in frequency and precoder index modulation techniques;
e.g., subcarrier index modulation in 
\cite{scim1},\cite{scim1a},\cite{scim2},\cite{scim2a},
and precoder index modulation in \cite{pim}. The focus of this paper is
twofold: $i)$ generalization of the idea of spatial modulation, which we
refer to as generalized spatial index modulation (GSIM), and $ii)$
generalization of the idea of index modulation to both spatial domain
(multiple-antennas) as well as frequency domain (subcarriers), which we
refer to as generalized space-frequency index modulation (GSFIM).

In spatial modulation, the choice of the transmit antenna to activate in
a channel use is made based on a group of $m$ bits, where the number of
transmit antennas is $n_t=2^m$. On the chosen antenna, a symbol from an
$M$-ary modulation alphabet $\mathbb{A}$ (e.g., $M$-QAM) is sent. The
remaining $n_t-1$ antennas remain silent. Therefore, the achieved rate in
spatial modulation, in bits per channel use (bpcu), is $\log_2n_t+\log_2M$.
The error performance of spatial modulation has been studied
extensively, and it has been shown that spatial modulation can achieve
performance gains compared to spatial multiplexing \cite{smp1},\cite{smp2}.
Space shift keying is a special case of spatial modulation \cite{ssk1},
where instead of sending an $M$-ary modulation symbol, a signal known to
the receiver, say +1, is sent on the chosen antenna. So, the achieved rate
in space shift keying is $\log_2 n_t$ bpcu. In spatial modulation and space
shift keying, the number of transmit RF chains is restricted one, and the
number of transmit antennas is restricted to powers of two. The first
contribution in this paper consists of generalization of spatial
modulation which removes these restrictions \cite{GSM1}-\cite{wcnc13},
an analysis of achievable rate, and proposal of a detection algorithm.
In generalized spatial index modulation (GSIM), the transmitter has
$n_t$ transmit antenna elements and $n_{rf}$ transmit RF chains,
$1\leq n_{rf}\leq n_t$, and $n_{rf}$ out of $n_t$ antennas are
activated at a time, thereby $\lfloor \log_2 {n_t\choose n_{rf}} \rfloor$
additional bits are conveyed through antenna indexing. Spatial modulation
and spatial multiplexing turn out to be as special cases of GSIM for
$n_{rf}=1$ and $n_{rf}=n_t$, respectively. We derive the expression for
the achievable rate in GSIM and easy-to-compute upper and lower bounds
on this rate. We show that the achievable rate in GSIM can be more than
that in spatial multiplexing, and analytically establish the condition
under which this can happen. It is noted that GSIM achieves this higher
rate using fewer transmit RF chains compared to spatial multiplexing.
We also propose a Gibbs sampling based detection algorithm for GSIM and
show that GSIM can achieve better bit error rate (BER) performance than
spatial multiplexing.

In the second contribution in this paper, we introduce GSFIM which uses
both spatial as well as frequency domain to encode bits through indexing.
GSFIM can be viewed as a generalization of the GSIM scheme
by exploiting indexing in the frequency domain as
well.  Index modulation that exploits the frequency domain alone --
referred to as subcarrier index modulation (SIM) -- has been studied in
\cite{scim1}-\cite{scim2a}. These works have shown that OFDM with
subcarrier index modulation (SIM-OFDM) achieves better performance than
conventional OFDM, particularly at medium to high SNRs. These works have
not exploited indexing in the spatial domain in MIMO systems. Our
contribution addresses, for the first time, indexing both in space as
well as frequency in MIMO systems. In particular, we $(i)$ propose a
signaling architecture for combined space and frequency indexing,
$(ii)$ study in detail its achieved rate in comparison with conventional
MIMO-OFDM, and $(iii)$ show that better performance compared to that in
conventional MIMO-OFDM can be achieved in the medium to high SNR regime.
The proposed GSFIM system has $N$ subcarriers, $n_t$ transmit antennas, and
$n_{rf}$ transmit RF chains, $1\leq n_{rf} \leq n_t$. In the spatial domain,
$n_{rf}$ out of $n_t$ transmit antennas are chosen for activation based
on $\lfloor \log_2{n_t\choose n_{rf}}\rfloor$ bits. In the frequency
domain, in a space-frequency block of size $n_{rf} \times N$, information
bits are encoded in multiple sub-blocks where each sub-block is of size
$n_{rf} \times n_f$ and $\frac{N}{n_f}$ is the number of sub-blocks.
We characterize the achievable rate in GSFIM as a function of the
system parameters. We show that GSFIM can offer better rates and less
transmit RF chains compared to those in conventional MIMO-OFDM. It is also
shown that GSFIM can achieve better BER performance than MIMO OFDM.

The rest of this paper is organized as follows. In Section \ref{gsm_1}, we
present the GSIM system model, and a detailed analysis of achievable rate
and rate bounds in GSIM. We quantify rate gains and savings in transmit RF
chains in GSIM compared to spatial multiplexing. The proposed detection
algorithm for GSIM and its BER performance are also presented. In Section
\ref{sec_gsfm}, we present the GSFIM system model, analysis of achievable
rate in GSFIM, and BER performance of GSFIM. Conclusions and scope for
future work are presented in Section \ref{sec_conc}.

\begin{figure}
\centering
\includegraphics[height=1.5in]{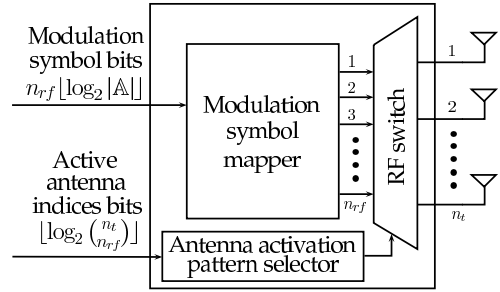}
\caption{GSIM transmitter.}
\label{fig_gsm_tx}
\vspace{-4mm}
\end{figure}

\vspace{-1mm}
\section{Generalized spatial index modulation}
\label{gsm_1}
In this section, we consider generalized spatial index modulation (GSIM)
which encodes bits through indexing in the spatial domain. In GSIM, the
transmitter has $n_t$ transmit antennas and $n_{rf}$ transmit RF chains,
$1\leq n_{rf}\leq n_t$. In any given channel use, $n_{rf}$ out of $n_t$
antennas are activated. Information bits are conveyed through both
conventional modulation symbols as well as the indices of the active
antennas. Spatial multiplexing becomes a special case of GSIM with
$n_{rf}=n_t$. We present an analysis of the achievable rates in GSIM,
which shows that the maximum achievable rate in GSIM can be more than the
rate in spatial multiplexing, and that too using fewer transmit RF chains.

\vspace{-1mm}
\subsection{System model}
A GSIM transmitter is shown in Fig. \ref{fig_gsm_tx}. It has $n_t$ transmit
antennas and $n_{rf}$ transmit RF chains, $1\leq n_{rf}\leq n_t$. An
$n_{rf}\times n_t$ switch connects the RF chains to the transmit antennas.
In a given channel use, $n_{rf}$ out of $n_t$ transmit antennas are chosen
and $n_{rf}$ $M$-ary modulation symbols are sent on these chosen antennas.
The remaining $n_t-n_{rf}$ antennas remain silent (i.e., they can be viewed
as transmitting the value zero). Therefore, if ${\mathbb A}$ denotes the
$M$-ary modulation alphabet used on the active antennas, the effective
alphabet becomes $\mathbb{A}_0 \Define \mathbb{A}\cup 0$.

Define an antenna activation pattern to be a $n_t$-length vector that
indicates which antennas are active (denoted by a `1' in the corresponding
antenna index) and which antennas are silent (denoted by a `0'). There are
$L = {n_t \choose n_{rf}}$ antenna activation patterns possible, and
$K = \big \lfloor \log_2{n_t \choose n_{rf}} \big \rfloor$ bits
are used to choose an activation pattern for a given channel use.
Note that not all $L$ activation patterns are needed, and any $2^K$
patterns out of them are adequate. Take any $2^K$ patterns out of $L$
patterns and form a set called the `antenna activation pattern set',
${\mathbb S}$. Let us illustrate this using the following example.
Let $n_t=4$ and $n_{rf}=2$. Then, $L={4 \choose 2}=6$,
$K = \left \lfloor \log_2 6 \right \rfloor = 2$, and $2^K=4$.
The six antenna activation patterns are given by

\vspace{2mm}
$\big\{[1,1,0,0]^T, [1,0,1,0]^T, [0,1,0,1]^T, [0,0,1,1]^T,$ 

\vspace{1mm}
\hspace{1mm} $[0,1,1,0]^T, [1,0,0,1]^T\big\}.$

\vspace{2mm}
Out of these six patterns, any $2^K=4$ patterns can be taken to form the
set ${\mathbb S}$. Accordingly, let us take the antenna activation pattern
set as
\[
{\mathbb S} = \big\{[1,1,0,0]^T, [1,0,1,0]^T, [0,1,0,1]^T, [0,0,1,1]^T\big\}.
\]
Table \ref{tab_gsim} shows the mapping of data bits to GSIM signals for
$n_t=4$, $n_{rf}=2$ for the above activation pattern set. Suppose 4-QAM
is used to send information on the active antennas. Let ${\bf x}\in
{\mathbb A}_0^{n_t}$ denote the $n_t$-length transmit vector. Let
$010011$ denote the information bit sequence. GSIM translates these
bits to the transmit vector ${\bf x}$ as follows: $i)$ the first two
bits are used to choose the activity pattern, $ii)$ the second two bits
form a 4-QAM symbol, and $iii)$ the third two bits form another 4-QAM
symbol, so that, with Gray mapping, the transmit vector ${\bf x}$ becomes
\[
{\bf x} \ = \ [1+{\bf j}, \ 0, \ -1-{\bf j}, \ 0]^T,
\]
where ${\bf j}=\sqrt{-1}$.

\begin{table}[t]
\centering
\begin{tabular}{|c|c|c|c|c|c| }
\hline
Data bits & Antenna activity
& \multicolumn{4}{|c|}{Antenna status} \\ [0.0ex] \cline{3-6}
\raisebox{-0.0ex}{$K=2$} & \raisebox{-0.0ex}{pattern} & \raisebox{-0.0ex}{Ant.1} & \raisebox{-0.0ex}{Ant.2}
& \raisebox{-0.0ex}{Ant.3} & \raisebox{-0.0ex}{Ant.4} \\ [0.0ex]
\hline \hline
\raisebox{-0.0ex}{$0 \ 0$} & \raisebox{-0.0ex}{$[1$, $1$, $0$, $0]^T$} & \raisebox{-0.0ex}{$\in {\mathbb A}$} & \raisebox{-0.0ex}{$\in {\mathbb A}$} & \raisebox{-0.0ex}{OFF} & \raisebox{-0.0ex}{OFF} \\[0ex]
\hline
\raisebox{-0.0ex}{$0 \ 1$} & \raisebox{-0.0ex}{$[1$, $0$, $1$, $0]^T$} & \raisebox{-0.0ex}{$\in {\mathbb A}$} & \raisebox{-0.0ex}{OFF} & \raisebox{-0.0ex}{$\in {\mathbb A}$} & \raisebox{-0.0ex}{OFF} \\ [0ex]
\hline
\raisebox{-0.0ex}{$1 \ 0$} & \raisebox{-0.0ex}{$[0$, $1$, $0$, $1]^T$} & \raisebox{-0.0ex}{OFF} & \raisebox{-0.0ex}{$\in {\mathbb A}$} & \raisebox{-0.0ex}{OFF} & \raisebox{-0.0ex}{$\in {\mathbb A}$} \\ [0ex]
\hline
\raisebox{-0.0ex}{$1 \ 1$} & \raisebox{-0.0ex}{$[0$, $0$, $1$, $1]^T$} & \raisebox{-0.0ex}{OFF} & \raisebox{-0.0ex}{OFF} & \raisebox{-0.0ex}{$\in {\mathbb A}$} & \raisebox{-0.0ex}{$\in {\mathbb A}$} \\ [0ex]
\hline
\end{tabular}
\caption[Data bits-to-GSIM signal mapping for $n_t=4$, $n_{rf}=2$.]{Data
bits to GSIM signal mapping for $n_t=4$, $n_{rf}=2$. \\ ${\mathbb A}$:
$M$-ary modulation alphabet. }
\vspace{-8mm}
\label{tab_gsim}
\end{table}

\vspace{-2mm}
\subsection{Achievable rates in GSIM}
\label{gsm_rate}
\begin{figure}
\centering
\includegraphics[height=2.5in, width=3.00in]{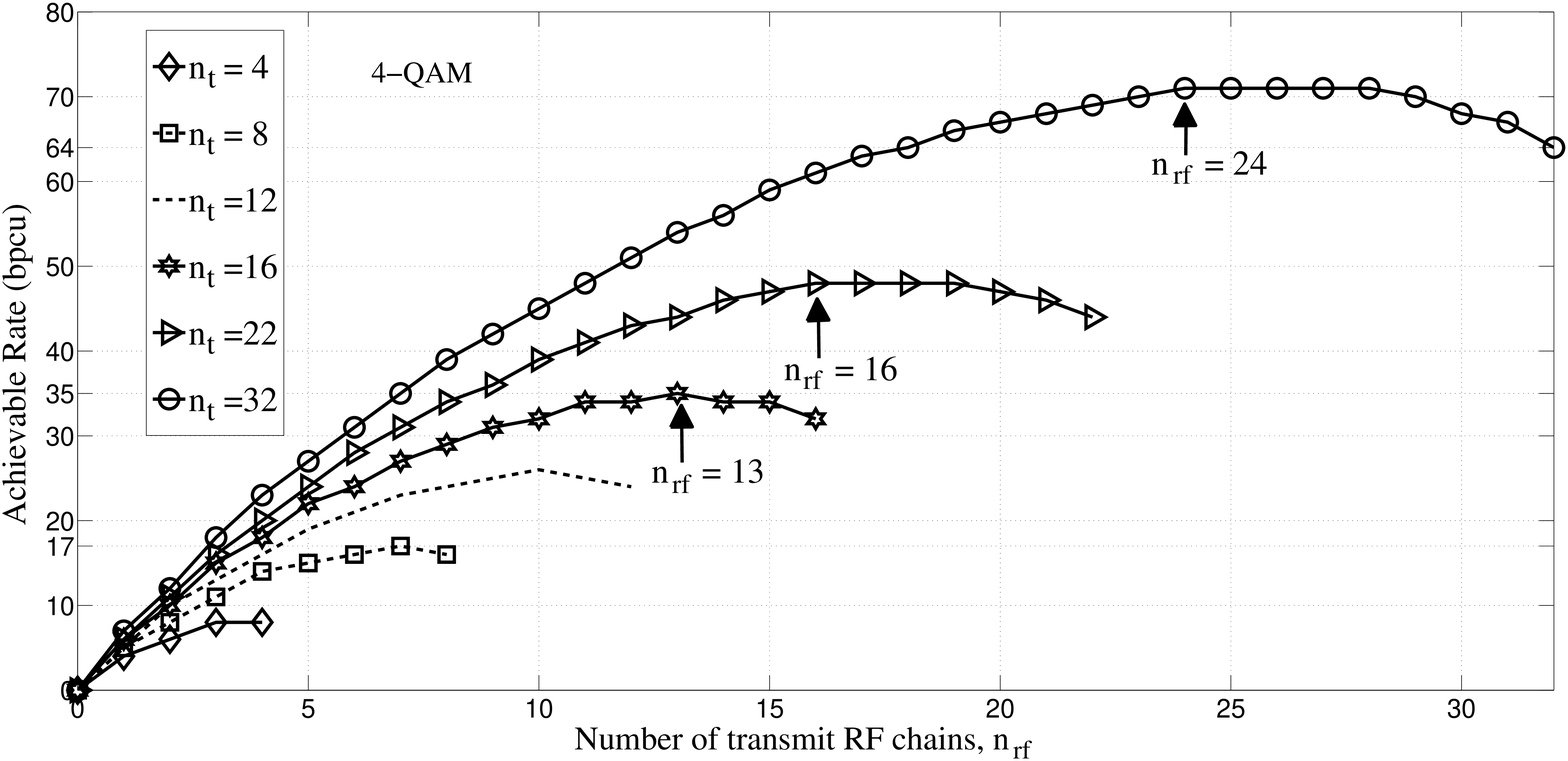}
\caption{Achievable rate in GSIM, $R_{\mbox{\scriptsize gsim}}$, as a
function of $n_{rf}$ for different values of $n_t$, and 4-QAM.}
\label{fig_gsm1}
\vspace{-4mm}
\end{figure}

The transmit vector in a given channel use in GSIM is formed using $i)$
antenna activation pattern selection bits, and $ii)$ $M$-ary modulation
bits. The number of activation pattern selection bits is
$\big\lfloor \log_2{n_t \choose n_{rf}}\big\rfloor$. The number of
$M$-ary modulation bits is $n_{rf}\log_2M$. Combining these two parts,
the achievable rate in GSIM with $n_t$ transmit antennas, $n_{rf}$
transmit RF chains, and $M$-QAM is given by
\begin{eqnarray}
R_{\mbox{\scriptsize gsim}} = \underbrace{\bigg\lfloor \log_2{n_t \choose n_{rf}}\bigg\rfloor}_{\mbox{\tiny Antenna index bits}} \ + \underbrace{n_{rf}\log_2M}_{\mbox{\tiny modulation symbol bits}} \quad \mbox{bpcu}.
\label{gsm_rate1}
\end{eqnarray}
Let us examine the GSIM rate $R_{\mbox{\scriptsize gsim}}$ in (\ref{gsm_rate1})
in some detail. In particular, let us examine how $R_{\mbox{\scriptsize gsim}}$
varies as a function of its variables. Fig. \ref{fig_gsm1} shows the
variation of $R_{\mbox{\scriptsize gsim}}$ as a function of $n_{rf}$ for
different values of $n_t=4,8,12,16,22,32$, and 4-QAM. The value of $n_{rf}$ in
the x-axis is varied from from 0 to $n_t$. As mentioned before, $n_{rf}=n_t$
corresponds to spatial multiplexing. The $R_{\mbox{\scriptsize gsim}}$ versus
$n_{rf}$ plot for a given $n_t$ shows an interesting behavior, namely,
for a given $n_t$, there is an optimum $n_{rf}$ that maximizes the
achievable rate $R_{\mbox{\scriptsize gsim}}$. Let 
$R_{\mbox{\scriptsize gsim}}^{max}$ denote the maximum achievable rate,
i.e.,
\begin{eqnarray}
R_{\mbox{\scriptsize gsim}}^{max} = \max_{1\leq n_{rf} \leq n_t}
R_{\mbox{\scriptsize gsim}}.
\label{gsm_ratemax}
\end{eqnarray}
In Fig. \ref{fig_gsm1}, it is interesting to see that
$R_{\mbox{\scriptsize gsim}}^{max}$ does not necessarily occur at
$n_{rf}=n_t$, but at some $n_{rf}<n_t$.
$R_{\mbox{\scriptsize gsim}}$ can exceed the spatial multiplexing rate of
$n_t\log_2M$ whenever the first term in (\ref{gsm_rate1}) exceeds
$(n_t-n_{rf})\log_2M$. The following theorem formally establishes the
condition under which the $R_{\mbox{\scriptsize gsim}}^{max}$ will be more
than the spatial multiplexing rate of $n_t\log_2M$.

\begin{thm}
The maximum achievable rate in GSIM is strictly greater than the
rate achieved in spatial multiplexing (i.e.,
$R_{\mbox{\scriptsize gsim}}^{max}>n_t\log_2M$)
iff $n_t\geq 2M$.
\end{thm}

{\bfseries {\em Proof:}}
Consider the two terms on the right-hand side (RHS) of the rate expression
(\ref{gsm_rate1}). The first term (contribution due to antenna index bits)
increases when $n_{rf}$ is increased from 0 to $\lfloor\frac{n_t}{2} \rfloor$
and then decreases, i.e., it peaks at $n_{rf}=\lfloor\frac{n_t}{2}\rfloor$.
The second term (contribution due to modulation symbol bits), on the other
hand, increases linearly with $n_{rf}$. These two terms when added can
cause a peak at some $n_{rf}$ in the range
$\lfloor\frac{n_t}{2}\rfloor \leq n_{rf} \leq n_t$.
Observe that, as we reduce $n_{rf}$ below $n_t$, we gain rate from the
first term but lose rate in the second term. The rate loss in the second
term is $\log_2M$ bpcu per RF chain reduced. Therefore, we can rewrite
(\ref{gsm_rate1}) as
\begin{eqnarray}
R_{\mbox{\scriptsize gsim}}
& = & n_t\log_2M + \left\lfloor \log_2{n_t \choose n_{rf}}\right\rfloor \nonumber \\
& & - (n_t-n_{rf})\log_2M.
\label{gsm_rate2a}
\end{eqnarray}

{\em Case 1:} $n_t\geq 2M$

If $n_t\geq 2M$, then $\left\lfloor\log_2 n_t\right\rfloor > \log_2M$.
By putting $n_{rf}=n_t-1$ in (\ref{gsm_rate2a}), we get
\begin{eqnarray}
R_{\mbox{\scriptsize gsim}}
&=& n_t\log_2M + \left\lfloor\log_2 n_t\right\rfloor - \log_2M.
\label{ntm1a}
\end{eqnarray}
Therefore, in this case, the $R_{\mbox{\scriptsize gsim}}$ in (\ref{ntm1a})
is more than $n_t\log_2M$, i.e., GSIM with
$n_{rf}=n_t-1$ RF chains achieves more rate than spatial multiplexing.
This implies $R_{\mbox{\scriptsize gsim}}^{max} > n_t\log_2M$, i.e., the
maximum rate available in GSIM is more than the spatial multiplexing rate.
Conversely, if $n_t< 2M$, we show below that
$R_{\mbox{\scriptsize gsim}}^{max}$
is not more than the spatial multiplexing rate.

{\em Case 2:} $n_t < 2M$

If $n_t< 2M$,
\begin{eqnarray}
\log_2 n_t&<&1+\log_2M.
\label{ineq1a}
\end{eqnarray}
From the properties of binomial coefficients, we have
\begin{eqnarray}
\hspace{-5mm}
{n_t\choose n_{rf}} &\hspace{-2mm} = &\hspace{-2mm} {n_t\choose {n_t- n_{rf}}} \nonumber \\
&\hspace{-2mm} = &\hspace{-2mm} \frac{n_t(n_t-1)\cdots(n_{rf}+1)}{1.2.\cdots (n_t-n_{rf})}
\ < \ \frac{n_t^{n_t-n_{rf}}}{2^{n_t-n_{rf}-1}}.
\label{ineq2a}
\end{eqnarray}
Hence,

\vspace{-2mm}
{\small
\begin{eqnarray}
\hspace{-5mm}
\left\lfloor\log_2 {n_t\choose n_{rf}}\right\rfloor & \hspace{-2.5mm} \leq & \hspace{-2.5mm}
\left\lfloor(n_t-n_{rf})\log_2n_t-n_t+n_{rf}+1\right\rfloor \label{ineq3a} \\
&\hspace{-28mm} < & \hspace{-15mm} \left\lfloor (n_t-n_{rf})(1+\log_2M)-n_t+n_{rf}+1\right\rfloor \label{ineq4a}\\
&\hspace{-28mm} < & \hspace{-15mm} (n_t-n_{rf}+1)\log_2M
\label{ineq5a}\\
&\hspace{-28mm} \leq & \hspace{-15mm} (n_t-n_{rf})\log_2M.
\label{ineq6a}
\end{eqnarray}
}

\vspace{-4mm}
\hspace{-5mm}
The inequality in (\ref{ineq3a}) is obtained by taking logarithm in
(\ref{ineq2a}), and  (\ref{ineq4a}) is obtained from (\ref{ineq3a})
and (\ref{ineq1a}). Hence, using (\ref{gsm_rate2a}), we obtain
$R_{\mbox{\scriptsize gsim}} \leq n_t\log_2M$, for $1\leq n_{rf}\leq n_t$,
and thus, for $n_t<2M$, $R_{\mbox{\scriptsize gsim}}^{max} \leq n_t\log_2M$.
Combining the arguments in Cases 1 and 2, we get {\bf Theorem 1}.
$\square$

From Fig. \ref{gsm_rate1}, the following interesting observations can
be made:
\begin{enumerate}
\item   by choosing the optimum $(n_t,n_{rf})$ combination (i.e., using
fewer RF chains than transmit antennas, $n_{rf}<n_t$), GSIM can achieve
a higher rate than that of spatial multiplexing where $n_{rf}=n_t$;
and
\item   one can operate GSIM at the same rate as that of spatial
multiplexing but with even fewer RF chains.
\end{enumerate}
For example, for $n_t=32$, the optimum $n_{rf}$ that maximizes
$R_{\mbox{\scriptsize gsim}}$ is 24 and the corresponding maximum rate,
$R_{\mbox{\scriptsize gsim}}^{max}$, is 71 bpcu. Compare this rate with
$32\log_24=64$ bpcu which is the rate achieved in spatial multiplexing.
This is a 11\% gain in rate in GSIM compared to
spatial multiplexing. Interestingly, this rate gain is achieved using
lesser number of RF chains; 24 RF chains in GSIM versus 32 RF chains
in spatial multiplexing. This is a 25\% savings in transmit RF chains
in GSIM compared to spatial multiplexing. Further, if GSIM were to
achieve the spatial multiplexing rate of 64 bpcu in
this case, then it can achieve it with even fewer RF chains, i.e.,
using just 18 RF chains which is a 43\% savings in RF chains compared
to spatial multiplexing. Table \ref{gsim_tab1} gives the percentage
gains in number of transmit RF chains at achieved rate
$R=R_{\mbox{\scriptsize gsim}}^{max}$ and $R=n_t\log_2M$, and the
percentage gains in rates achieved by GSIM compared to spatial
multiplexing for $n_t=16,32$ with BPSK, 4-QAM, 8-QAM, and 16-QAM.

\begin{table*}
\center
\begin{tabular}{|c||c|c||c|c||c|c|}
\hline
\multicolumn{1}{|p{1.2cm}||}
{$M$-ary \newline alphabet} & \multicolumn{2}{|p{3.7cm}||}{{\color{black}Percentage} saving in no. of Tx RF chains at {$R=R_{\mbox{\scriptsize gsim}}^{max}$}} & \multicolumn{2}{|p{4.0cm}||}{{\color{black}Percentage} saving in no. of Tx RF chains at {$R=n_t\log_2M$}} & \multicolumn{2}{|p{3.5cm}|} {{\color{black}Percentage} increase in rate at  $R=R_{\mbox{\scriptsize gsim}}^{max}$} \\ \cline{2-7}
& $n_t=16$ & $n_t=32$ & $n_t=16$ & $n_t=32$ & $n_t=16$ & $n_t=32$ \\
\hline \hline
BPSK & 31.25 & 40.63 & 68.75 & 71.88 & 43.75 & 46.88 \\ \hline
4-QAM & 18.75 & 25 & 37.5 & 43.75 & 9.385 & 10.94 \\ \hline
8-QAM & 6.25 & 12.5 & 18.75 & 21.88 & 2.08 & 3.13 \\ \hline
16-QAM & 6.25 & 3.13 & 6.25 & 9.38 & 0 & 0.78 \\ \hline
\end{tabular}
\caption{Percentage saving in transmit RF chains and
percentage increase in rate in GSIM compared to spatial
multiplexing for $n_t=16,32$ and BPSK, 4-/8-/16-QAM.}
\vspace{-6mm}
\label{gsim_tab1}
\end{table*}

\subsection{Bounds on  achievable rates in GSIM}
\label{gsm_bnd}
We now proceed to obtain bounds on the achievable rate in GSIM.
From (\ref{gsm_rate1}), we observe that
\begin{eqnarray}
R_{\mbox{\scriptsize gsim}} & \hspace{-2mm} \leq & \hspace{-2mm} \log_2\left(\frac{n_t!}{n_{rf}!(n_t-n_{rf})!}\right)+n_{rf}\log_2M,
\label{rate2}
\end{eqnarray}
and
\begin{eqnarray}
\hspace{-4mm}
R{\mbox{\scriptsize gsim}} &\hspace{-2mm} > & \hspace{-2mm} \log_2\left(\frac{n_t!}{n_{rf}!(n_t-n_{rf})!}\right)+n_{rf}\log_2M-1.
\label{rate3}
\end{eqnarray}
From the properties of the factorial operator \cite{wikistirling}, we have
\begin{eqnarray}
\sqrt{2\pi n}\left(\frac{n}{e}\right)^n \ \leq \ n! \ \leq \ e\sqrt n\left(\frac{n}{e}\right)^n, \quad \forall n\in \mathbb{N}.
\label{factbnd}
\end{eqnarray}
Let us define the function $f(n_t,n_{rf},\log_2 {M})$ as
\begin{eqnarray}
\hspace{-4mm}
f(n_t,n_{rf},\log_2 M) & \hspace{-2mm} \Define & \hspace{-2mm} n_t\log_2 n_t-n_{rf}\log_2 n_{rf} \nonumber \\
&\hspace{-33mm} & \hspace{-27mm} - (n_t-n_{rf})\log_2 (n_t-n_{rf})+n_{rf}\log_2 M.
\label{fnr}
\end{eqnarray}
Substituting (\ref{factbnd}) in (\ref{rate2}), using (\ref{fnr}), and
simplifying, we get
\begin{eqnarray}
R_{\mbox{\scriptsize gsim}} & \hspace{-2mm} \leq & \hspace{-2mm} \log_2 \frac{e}{2\pi}+0.5\log_2\frac{n_t}{n_{rf}(n_t-n_{rf})} \nonumber \\
& \hspace{-2mm} & \hspace{-2mm} +f(n_t,n_{rf},\log_2M).
\label{rbnd1}
\end{eqnarray}
In a similar way, using (\ref{factbnd}) in (\ref{rate3}), we can write
\begin{eqnarray}
R_{\mbox{\scriptsize gsim}} & \hspace{-2mm} > & \hspace{-2mm} \log_2 \frac{\sqrt{2\pi}}{e^2} + 0.5\log_2\frac{n_t}{n_{rf}(n_t-n_{rf})} \nonumber \\
& \hspace{-2mm} & \hspace{-2mm} +f(n_t,n_{rf},\log_2M)-1.
\label{rbnd2}
\end{eqnarray}
Let us rewrite (\ref{rbnd1}) and (\ref{rbnd2}) in the following way:
\begin{eqnarray}
R_{gsim} & \leq & f_1(n_t, n_{rf})+f_2(n_t, n_{rf})+c_1,
\end{eqnarray}
and
\begin{eqnarray}
R_{gsim} & > & f_1(n_t, n_{rf})+f_2(n_t, n_{rf})+c_2,
\end{eqnarray}
where
$f_1(n_t, n_{rf}) = 0.5\log_2\frac{n_t}{n_{rf}(n_t-n_{rf})}$,
$f_2(n_t, n_{rf})  = f(n_t, n_{rf}, \log_2{M})$,
$c_1  = \log_2\frac{e}{2\pi}$, and
$c_2  = \log_2 \frac{\sqrt{2\pi}}{e^2}-1$.
For a fixed $n_t$, the maximum value of $f_1(n_t, n_{rf})$ in the
range $1\leq n_{rf} \leq n_t-1$ is obtained at $n_{rf}=1$ or $n_{rf}=n_t-1$,
and the maximum value is $0.5\log_2\big(\frac{n_t}{n_t-1}\big)$.
Hence,
\begin{eqnarray}
\max\{f_1(n_t, n_{rf})\}=  0.5\log_2\frac{n_t}{n_t-1}.
\label{newx1}
\end{eqnarray}
Also, the term $f_1(n_t, n_{rf})$ is minimized for
$n_{rf}=\lfloor\frac{n_t}{2}\rfloor$, and the minimum value is
$0.5\log_2\frac{4}{n_t}=1-0.5\log_2 n_t$ for even $n_t$, and is
$0.5\log_2\frac{n_t}{(\frac{n_t}{2})^2-0.25}\geq
1-0.5\log_2 n_t$ for odd $n_t$. Hence,
\begin{eqnarray}
\min\{f_1(n_t, n_{rf})\} \geq 1-0.5\log_2 n_t .
\label{newx2}
\end{eqnarray}
Therefore, from (\ref{newx1}), (\ref{newx2}) and (\ref{rbnd1}), (\ref{fnr}),
we obtain the upper bound on $R_{\mbox{\scriptsize gsim}}$ as
\begin{eqnarray}
R_{\mbox{\scriptsize gsim}} & \hspace{-2mm} \leq & \hspace{-2mm} f(n_t,n_{rf},\log_2M)+0.5\log_2\frac{n_t}{n_t-1} \nonumber \\
&\hspace{-2mm} & \hspace{-2mm} +\log_2 \frac{e}{2\pi}.
\label{Rmaxbnd01}
\end{eqnarray}
In a similar way, from (\ref{rbnd2}) and (\ref{fnr}), we obtain the lower
bound on $R_{\mbox{\scriptsize gsim}}$ as
\begin{eqnarray}
R_{\mbox{\scriptsize gsim}} & \hspace{-2mm} > &\hspace{-2mm} f(n_t,n_{rf},\log_2M)-0.5\log_2 n_t \nonumber \\
& \hspace{-2mm} & \hspace{-2mm} + \log_2 \frac{\sqrt{2\pi}}{e^2}.
\label{Rmaxbnd02}
\end{eqnarray}
Since $n_t$, $n_{rf}$ and $M$ take finite positive integer values,
and because of the floor operation in the first term on the RHS in
(\ref{gsm_rate1}),
we can rewrite the bounds in (\ref{Rmaxbnd01}) and (\ref{Rmaxbnd02}) as
\begin{eqnarray}
R_{\mbox{\scriptsize gsim}} & \hspace{-2mm} \leq & \hspace{-2mm} \bigg\lfloor f(n_t,n_{rf},\log_2M)+0.5\log_2\frac{n_t}{n_t-1} \nonumber \\
& \hspace{-2mm} & \hspace{-2mm} +\log_2 \frac{e}{2\pi}\bigg\rfloor,
\label{Rmaxbnd03}
\end{eqnarray}
and
\begin{eqnarray}
R_{\mbox{\scriptsize gsim}} & \hspace{-2mm} \geq & \hspace{-2mm} \bigg\lceil f(n_t,n_{rf},\log_2M)-0.5\log_2 n_t \nonumber \\
& \hspace{-2mm} & \hspace{-2mm} +\log_2 \frac{\sqrt{2\pi}}{e^2}\bigg\rceil.
\label{Rmaxbnd04}
\end{eqnarray}
Note that the above bounds on $R_{\mbox{\scriptsize gsim}}$ can be
computed easily for any $n_t$, $n_{rf}$, without the need for the
computation of factorials of large numbers in the actual rate
expression in (\ref{gsm_rate1}). 
Further, noting that the optimum $n_{rf}$ that maximizes $f_2(n_t,n_{rf})$
is given by
\begin{eqnarray}
n_{rf}^* & = & \frac{n_t{M}}{{M}+1},
\label{rstar}
\end{eqnarray}
we obtain upper and lower bounds on $R_{\mbox{\scriptsize gsim}}^{max}$,
by substituting $n_{rf}^*$ in (\ref{rstar}) into (\ref{Rmaxbnd03}) and
(\ref{Rmaxbnd04}), respectively, as
\begin{eqnarray}
R_{\mbox{\scriptsize gsim}}^{max} &\hspace{-2mm} \leq & \hspace{-2mm} \bigg\lfloor n_t\log_2 ({M}+1)+0.5\log_2\frac{n_t}{n_t-1} \nonumber \\
& \hspace{-2mm} & \hspace{-2mm} +\log_2 \frac{e}{2\pi}\bigg\rfloor,
\label{Rmaxbnd5}
\end{eqnarray}
and
\begin{eqnarray}
R_{\mbox{\scriptsize gsim}}^{max} & \hspace{-2mm} \geq & \hspace{-2mm} \bigg\lceil f\left(n_t,\left\lfloor n_t\frac{{M}}{{M}+1}\right\rceil,\log_2M\right)-0.5\log_2 n_t \nonumber \\
& \hspace{-2mm} & \hspace{-2mm} +\log_2 \frac{\sqrt{2\pi}}{e^2}\bigg\rceil.
\label{Rmaxbnd6}
\end{eqnarray}
These bounds on $R_{\mbox{\scriptsize gsim}}^{max}$ can be calculated for any
given $n_t$ and $M$ directly, without exhaustive computation of the rate
for all possible values of $n_{rf}$.
From  (\ref{Rmaxbnd5}) and (\ref{Rmaxbnd6}), we observe that as
$n_t\rightarrow \infty$, $R_{\mbox{\scriptsize gsim}}^{max}$ can be
approximated by $n_t\log_2(M+1)$.
Note that a spatial
multiplexing system which uses a zero-augmented alphabet $\mathbb{A}_0$
achieves the rate of $n_t\log_2 (M+1)$,
if all the symbols
in $\mathbb{A}_0$ are equiprobable.

\begin{figure}
\centering
\subfigure[]{
\includegraphics[width=2.75in,height=2.10in]{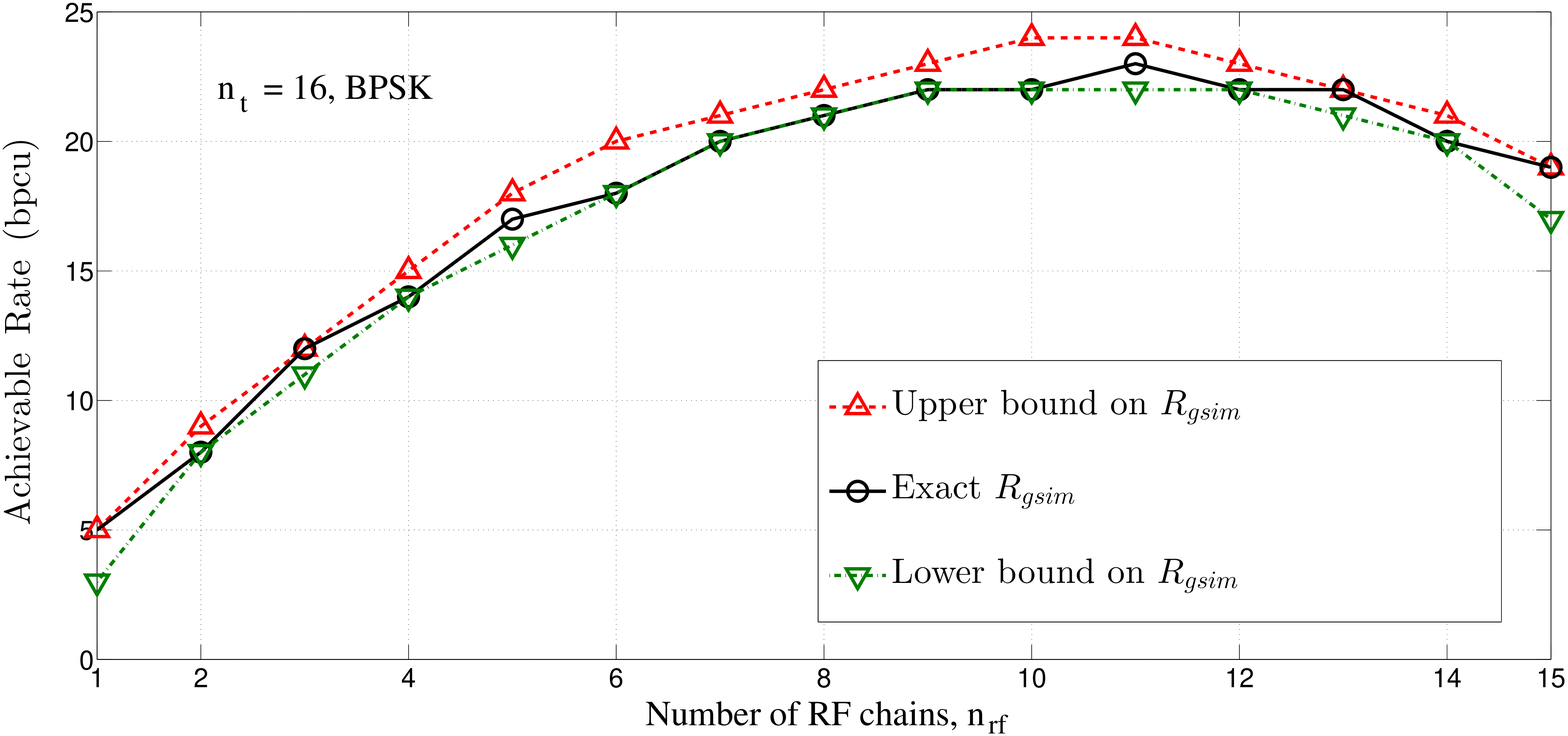}
\label{fig6_2a}
}
\hspace{5mm}
\subfigure[]{
\includegraphics[width=2.75in,height=2.10in]{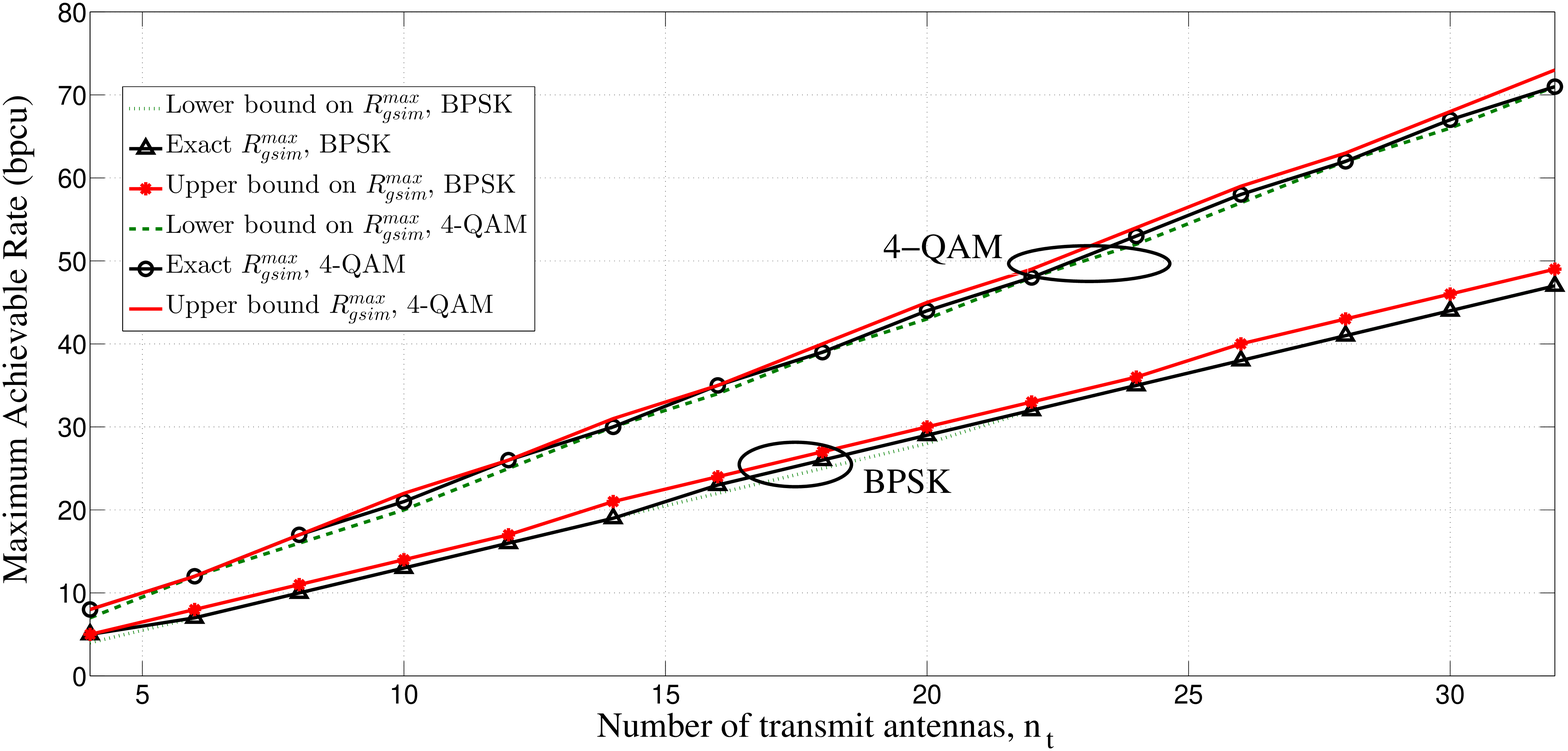}
\label{fig6_3}
}
\caption{(a) Bounds on $R_{\mbox{\scriptsize gsim}}$ with BPSK for $n_t=16$
and varying $n_{rf}$. (b) Bounds on $R_{\mbox{\scriptsize gsim}}^{max}$
with BPSK and 4-QAM for varying $n_t$.}
\vspace{-4mm}
\end{figure}

In Fig. \ref{fig6_2a}, we plot the upper and lower bounds of
$R_{\mbox{\scriptsize gsim}}$ computed using (\ref{Rmaxbnd03}) and
(\ref{Rmaxbnd04}), respectively, along with exact
$R_{\mbox{\scriptsize gsim}}$, for $n_t=16$ and BPSK ($M=2$).
The number of RF chains, $n_{rf}$, is varied from 1 to 15. It can be
observed that the upper and lower bounds are tight (within $2$ bpcu of
the actual rate). In Fig. \ref{fig6_3}, we plot the upper and lower bounds
of $R_{\mbox{\scriptsize gsim}}^{max}$ obtained from (\ref{Rmaxbnd5}) and
(\ref{Rmaxbnd6}), respectively, for different values of $n_t$ and $M=2,4$
(i.e., BPSK, 4-QAM). The corresponding exact 
$R_{\mbox{\scriptsize gsim}}^{max}$ values are also plotted for 
comparison. It can be observed that the lower and upper bounds of 
$R_{\mbox{\scriptsize gsim}}^{max}$ are within 
$2$ bpcu of the exact $R_{\mbox{\scriptsize gsim}}^{max}$.

\subsection{GSIM signal detection}
\label{gsm_3}
In this subsection, we consider detection of GSIM signals.
Let ${\bf H}$ denote the $n_r\times n_t$ channel matrix, where $n_r$ is
the number of receive antennas. Assume rich scattering environment where
the entries of ${\bf H}$ are modeled as circularly symmetric complex
Gaussian with zero mean and unit variance. Let ${\bf y}$ denote the
$n_r\times 1$-sized received vector, which is given by
\begin{eqnarray}
{\bf y}={\bf H}{\bf x}+{\bf n},
\label{sys1}
\end{eqnarray}
where ${\bf x}$ is the $n_t\times 1$-sized transmit vector and ${\bf n}$
is the $n_r\times 1$-sized additive white Gaussian noise vector at the
receiver, whose $i$th element $n_i\sim \mathcal{CN}(0,\sigma^2)$,
$\forall i=1,2,\cdots,n_r$.
Let ${\mathbb U}$ denote the set of all possible transmit vectors,
given by
\begin{eqnarray}
\mathbb{U}& = & \{{\bf x} | {\bf x}\in {\mathbb{A}_0}^{n_t\times 1},\|{\bf x}\|_0=n_{rf},{\bf t}^{\bf x}\in \mathbb{S}\},
\label{Udef}
\end{eqnarray}
where $\|{\bf x}\|_0$ denotes the zero norm of vector ${\bf x}$ (i.e.,
number of non-zero entries in ${\bf x}$), and ${\bf t}^{\bf x}$ denotes
the antenna activation pattern vector corresponding to ${\bf x}$, where
$t^{\bf x}_j=1, \text{iff} \, \, {x}_j\neq 0, \forall j=1,2,\cdots,n_t.$
Note that $|\mathbb{U}|=2^{R_{\mbox{\scriptsize gsim}}}$.
The activation pattern set $\mathbb{S}$ and the mapping between elements of
$\mathbb{S}$ and antenna selection bits are known at both transmitter and
receiver. Hence, from (\ref{sys1}) and (\ref{Udef}), the ML decision rule
for GSIM signal detection is given by
\begin{eqnarray}
\widehat{{\bf x}} & = & \arg\min_{{\bf x} \in \mathbb{U}}\|{\bf y}-{\bf H}{\bf x} \|^2.
\label{eq_gsm_ml}
\end{eqnarray}
For small values of $n_t$ and $n_{rf}$, the set $\mathbb{U}$ may be fully
enumerated and ML detection as per (\ref{eq_gsm_ml}) can be done. But for
medium and large values of $n_t$ and $n_{rf}$, brute force computation of
$\widehat{{\bf x}}$ in (\ref{eq_gsm_ml}) becomes computationally prohibitive.
Here, we propose a low complexity algorithm for detection of GSIM signals.

The proposed approach is based on Gibbs sampling, where a Markov chain is
formed with all possible transmitted vectors as states. As the total number
of non-zero entries in the solution vector has to be equal to $n_{rf}$,
one can not sample each coordinate individually as is done in the case of
Gibbs sampling based detection in conventional MIMO systems \cite{mcmc}.
To address this issue, we propose the following sampling approach: 
sample two coordinates at a time jointly, keeping other $(n_t-2)$
coordinates fixed which contain $(n_{rf}-1)$ non-zero entries.

\subsubsection{Proposed modified Gibbs sampler}
For any vector
${\bf x}^{(t)}\in {\mathbb A}_0^{n_t}, \|{\bf x}^{(t)}\|_0={n_{rf}}$,
where the $t$ in the superscript of ${\bf x}^{(t)}$ refers to the
iteration index in the algorithm. Let $i_1,i_2,\cdots, i_{n_{rf}}$
denote the locations of non-zero entries and
$j_1,j_2,\cdots,j_{n_t-n_{rf}}$ denote the locations of zero entries
in ${\bf x}^{(t)}$. We will sample ${ x}_{i_l}^{(t)}$ and
${x}_{j_k}^{(t)}$ jointly, keeping other coordinates fixed, where
$l=1,2,\cdots,n_{rf}$ and $k=1,2,\cdots,(n_t-n_{rf})$.
As any possible transmitted vector can have only $n_{rf}$ non-zero
entries, the next possible state ${\bf x}^{(t+1)}$ can only be any
one of the following $2|\mathbb{A}|$ candidate vectors denoted by
$\{{\bf z}^w, w= 1,2,\cdots, 2|\mathbb{A}|\}$, which can be
partitioned into two sets. In the first set corresponding to
$w=1,2,\cdots,|\mathbb{A}|$, we enlist the vectors which have the
same activity pattern as ${\bf x}^{(t)}$. Hence,
${z}^w_{i_l}=\mathbb{A}^w, z^w_{j_k}=0, z_q=x^{(t)}_q, q=1,2,\cdots,n_t, q\neq i_l, j_k, \forall w=1,2,\cdots,|\mathbb{A}|$.
For $w= |\mathbb{A}|+1,|\mathbb{A}|+2,\cdots, 2|\mathbb{A}|$, we enlist
the vectors whose activity pattern differs from that of ${\bf x}^{(t)}$
in locations $j_k$ and $i_l$. Hence,
${z}^w_{i_l}=\mathbb{A}^w, z^w_{j_k}=0, z_q=x^{(t)}_q, q=1,2,\cdots,n_t, q\neq i_l, j_k, \forall w=1,2,\cdots,|\mathbb{A}|$.
For $w= |\mathbb{A}|+1,|\mathbb{A}|+2,\cdots, 2|\mathbb{A}|$, we enlist
the vectors whose activity pattern differs from that of ${\bf x}^{(t)}$
in locations $j_k$ and $i_l$. Hence,
${z}^w_{j_k}= \mathbb{A}^{(w-|\mathbb{A}|)}, z^w_{i_l}=0, z_q=x^{(t)}_q, q=1,2,\cdots,n_t, q\neq i_l, j_k, \forall w=|\mathbb{A}|+1,|\mathbb{A}|+2,\cdots,2|\mathbb{A}|$.

To simplify the sampling process, we calculate the best vectors from the
two sets corresponding to not swapping and swapping the zero and non-zero
locations, and choose among these two vectors. Let ${\bf x}^{NS}$ denote
the best vector from the first set corresponding to no swap. We set
${\bf x}^{NS}={\bf x}^{(t)}+ \lambda {\bf e}_{i_l}$ and minimize
$\|{\bf y}- {\bf H}{{\bf x}^{NS}}\|^2$ over $\lambda$. For this, we have
\begin{eqnarray}
\|{\bf y}- {\bf H}{{\bf x}^{NS}}\|^2 &\hspace{-2mm} = & \hspace{-2mm} \|{\bf y}- {\bf H}({\bf x}^{(t)}+ \lambda {\bf e}_{i_l})\|^2 \nonumber \\
&\hspace{-2mm}=&\hspace{-2mm} {\bf y}^H {\bf y}-2\Re\left({\bf y}^{MF}{\bf x}^{(t)}\right)+{{\bf x}^{(t)}}^H{\bf R}{\bf x}^{(t)} \nonumber \\
&\hspace{-35mm} & \hspace{-28mm} -\:2\Re\left(\lambda{\bf y}^{MF} {\bf e}_{i_l}\right)+2\Re\left(\lambda{{\bf x}^{(t)}}^H{\bf R}{\bf e}_{i_l}\right)+|\lambda|^2 R_{i_l,i_l},
\label{diff}
\end{eqnarray}
where ${\bf y}^{MF}={\bf y}^H{\bf H}$ and ${\bf R}={\bf H}^H{\bf H}$.
Differentiating (\ref{diff}) w.r.t $\lambda$ and equating it to zero, we get
\begin{eqnarray}
\lambda_{opt} & = & \frac{\left({y^{MF}_{i_l}-{{\bf x}^{(t)}}}^H {\bf r}_{i_l}\right)^H}{ R_{i_l,i_l}},
\label{lambdaopt}
\end{eqnarray}
where ${\bf r}_{i_l}$ is the $i_l$th column vector of $\bf R$. We obtain
${\bf x}^{NS}=[{\bf x}^{(t)}+\lambda_{opt}{\bf e}_{i_l}]_\mathbb{A}$, where
$[{\bf x}]_\mathbb{A}$ denotes the element-wise quantization of ${\bf x}$
to its nearest point in $\mathbb{A}$.
Similarly, we obtain ${\bf x}^S$, the best vector from the second set
corresponding to swap. The next state ${\bf x}^{(t+1)}$ is chosen between
${\bf x}^{S}$ and ${\bf x}^{NS}$ with probability $p^S$ and $p^{NS}$,
respectively, where ${p}^S=(1-q) \widetilde{p}^S+\frac{q}{2}$,
${p}^{NS}=1-{p}^S$, and
\begin{eqnarray}
\widetilde{p}^S&=&\frac{\exp(-\frac{\|{\bf y}- {\bf H}{{\bf x}^{S}}\|^2-\|{\bf y}- {\bf H}{{\bf x}^{NS}}\|^2}{\sigma^2})}{1+\exp(-\frac{\|{\bf y}- {\bf H}{{\bf x}^{S}}\|^2-\|{\bf y}- {\bf H}{{\bf x}^{NS}}\|^2}{\sigma^2})}.
\label{ps}
\end{eqnarray}
Here, $q$ gives the probability of mixing between Gibbs sampling and
sampling from uniform distribution. We use $q=\frac{1}{n_t}$,
because the simulation plots of BER as a function of $q$ have shown that
the best BER is achieved at around $q=\frac{1}{n_t}$.
After sampling, the best vector
obtained so far is updated. The above sampling process is repeated for
all $l$ and $k$. The algorithm is stopped after it meets the stopping
criterion or reaches the maximum number of allowable iterations, and
outputs the best vector in terms of ML cost obtained so far.

\subsubsection{Stopping and restart criterion}
The following stopping criterion and restart criterion are employed
in the algorithm. Let us denote the best vector so far as ${\bf z}$.
The stopping criterion works as follows: compute a metric
$\Theta_s({\bf z})=\Big\lceil \max\big(c_{min}, c_1\exp(\phi({{\bf z}}))\big)\Big\rceil$,
where
$\phi({\bf z})=\frac{\|{\bf y}-{\bf H}\hat{{\bf  x}}\|^2-n_r\sigma^{2}}{\sqrt{n_r}\sigma^{2}}$
is the normalized ML cost of ${\bf z}$. If ${\bf z}$ has not changed for
$\Theta_s({\bf z})$ iterations, then stop. This concludes one restart and
${\bf z}$ is declared as the output of this restart. Now, check whether
${\bf t}^{{\bf x}}$ belongs to $\mathbb{S}$ or not to check its validity.
Several such runs, each starting from a different initial vector, are
carried out till the best valid output obtained so far is reliable in
terms of ML cost. Let us denote the best vector among restart outputs
as ${\bf s}$ and the number of restarts that has given ${\bf s}$ as
output as $r_s$. We calculate another metric
{$\Theta_r({\bf s})=\left \lfloor \max\left(0,c_2\phi({{\bf s}})\right)\right\rfloor+1$}
and compare $r_s$ with this. If $r_s$ is equal to $\Theta_r({\bf s})$ or
maximum number of restarts is reached, we terminate the algorithm. The
listing of the proposed algorithm is given in {\bf Algorithm \ref{alg_gsm}}.

\begin{algorithm}
{
{\small 
\caption[Proposed Gibbs sampling based algorithm for GSIM detection.]{Proposed Gibbs sampling based algorithm for GSIM detection}
\label{alg_gsm}
\begin{algorithmic}[1]
\STATE {{\bf input:} ${\bf y}$, ${\bf H}$, $n_t, n_{rf};$
 {\small MAX-ITR: max. no. of iterations; MAX-RST: max. no. of restarts;}}
\STATE Compute ${\bf y}^{MF}={\bf y}^H{\bf H}$ and ${\bf R}={\bf H}^H{\bf H}$; initialize $r=0$, \,\, ${\kappa}=10^{10}$, \,\, $q=\frac{1}{n_t}$;
\STATE  $\phi(.):$ ML cost fn; \,\,
$ \Theta_s(.):$ \textit{stopping criterion} fn;
$ \Theta_r(.):$ \textit{restart criterion} fn;
\WHILE {$r <$ MAX-RST}
\STATE {${\bf{x}}^{(0)}:$ initial vector $\in {\mathbb{A}_0}^{n_t\times 1}; \|{\bf{x}}^{(0)}\|_0=n_{rf}$; $\beta=\phi({\bf{x}}^{(0)}); \quad {\bf z}={\bf x}^{(0)};\quad t=0$};\\
\WHILE{$t <$ MAX-ITR}
\FOR{$l=1$ to $n_{rf}$}
\FOR{$k=1$ to $n_t-n_{rf}$}
\STATE {find $i_l$ and $j_k$ indices;}\\
\STATE{ Compute $\lambda_{opt}$ from (\ref{lambdaopt})};\quad compute ${\bf x}^{NS}=[{\bf x}^{(t)}+\lambda_{opt}{\bf e}_{i_l}]_\mathbb{A}$;
compute ${\bf x}^{NS}$;\\
\STATE{ Compute $\widetilde{p}^S$ from (\ref{ps}); \,\,compute ${p}^S=(1-q) \widetilde{p}^S+\frac{q}{2}$,\,\, ${p}^{NS}=1-{p}^S$;}\\
\STATE{ Choose ${\bf x}^{(t+1)} $ between ${\bf x}^{S}$ and
${\bf x}^{NS}$ with probability $p^S$ \& $p^{NS}$;}
\STATE {$\gamma=\phi({\bf{x}}^{(t+1)});$}\\
\IF{$(\gamma \leq \beta)$}
\STATE ${\bf z}={\bf x}^{(t+1)};$ \,\,\, $\beta=\gamma$;\,\,\,calculate $\Theta_s({\bf z})$;\\
\ENDIF\\
\STATE {$t=t+1;$}\,\, $\beta_v^{(t)}=\beta;$\\
\ENDFOR\\
\ENDFOR\\
\IF{$\Theta_s({\bf z})<t$}
\IF {$\beta_v^{(t)}==\beta_v^{\left( t-\Theta_s({\bf z})\right)}$}
\STATE goto step 26 \\
\ENDIF\\
\ENDIF\\
\ENDWHILE\\
\STATE{ $r=r+1; \quad $}\\
\IF {${\bf t}^{{\bf z}}\in \mathbb{S}$}
\IF {$\beta<\kappa$}
\STATE{$\kappa=\beta$;\quad $r_{s}=1$; \quad ${\bf s}={\bf z}$; \quad Compute $\Theta_r({\bf s})$;}\\
\ENDIF \\
\IF{$\beta==\kappa$}
\STATE{$r_{s}=r_{s}+1;$}\\
\ENDIF\\
\IF {$r_{s}==\Theta_r({\bf s})$}
\STATE goto step 39\\
\ENDIF\\
\ENDIF\\
\ENDWHILE\\
\STATE {{\bf output:} ${\bf s}. \qquad{\bf s}:$ output solution vector}\\
\end{algorithmic}
}
}
\end{algorithm}

\subsubsection{Complexity}
The complexity of the proposed Gibbs sampling based detector can be
separated into three parts: $i$) computation of starting vectors, $ii$)
computation of ${\bf y}^{MF}$ and ${\bf R}$, and $iii$) computations
involved in the sampling and updating process. In our simulations, we
use MMSE output as the starting vector for the first restart, and
random starting vectors for the subsequent restarts. The MMSE output
needs the computation of
$\big({\bf H}^H{\bf H}+ \sigma^2{\bf I}_{n_t}\big)^{-1}{\bf H}^H{\bf y}$,
whose complexity is $\mathcal{O}(n_t^3)$. Note that this operation
includes the computations of ${\bf y}^{MF}$ and ${\bf R}$. For the
sampling and updating process, in each iteration, i.e., for each choice
of $l$ and $k$, the algorithm needs to compute
${{\bf x}^{(t)}}^H {\bf r}_{i_l}$ and ${{\bf x}^{(t)}}^H {\bf r}_{j_k}$,
which requires $\mathcal{O}(n_{rf})$ computations. The rest of the
computations are $\mathcal{O}(1)$. The number of iterations before the
algorithm terminates is found to be $\mathcal{O}(n_{rf}(n_t-n_{rf}))$
by computer simulations. Thus, the total number of computations involved
in $iii$) is $\mathcal{O}(n_{rf}^2(n_t-n_{rf}))$. Hence, the total
complexity of the proposed algorithm for GSIM detection is
$\mathcal{O}(n_t^3)+\mathcal{O}(n_{rf}^2(n_t-n_{rf}))$.

\subsection{BER performance results}
\label{sec6_6}
We now present the BER performance of GSIM. For systems with small $n_t$,
we present brute-force ML detection performance. For systems with large
$n_t$ where brute-force ML detection is prohibitive, we present the
performance using the proposed detection algorithm. We also compare
the performance of GSIM with the performance of spatial multiplexing.
For notation purpose, a GSIM system with $n_t$ transmit antennas and
$n_{rf}$ transmit RF chains is referred to as ``$(n_t,n_{rf})$-GSIM''
system. Also, we use the term ``$(n_t,n_{rf})$-SM'' system to refer
the spatial multiplexing system where $n_t=n_{rf}$. The following
parameters are used in proposed detection algorithm:
$c_{min}=10n_{rf}(n_t-n_{rf})$, $c_1=10n_{rf}(n_t-n_{rf})\log_2M$,
MAX-ITR = $8n_tn_{rf}(n_t-n_{rf})\sqrt{M}$, MAX-RST$=20$, $c_2=0.5(1+\log_2M)$.
Let $n_{rf}^{mid}$ denote the minimum number of RF chains in GSIM that
achieves the same rate as in spatial multiplexing for a given $n_t$
and $M$. Let $n_{rf}^{opt}$ denote the number of RF chains that
achieves $R_{\mbox{\scriptsize gsim}}^{max}$ for a given $n_t$ and $M$.

\begin{figure}
\centering
\includegraphics[height=2.5in,width=3.00in]{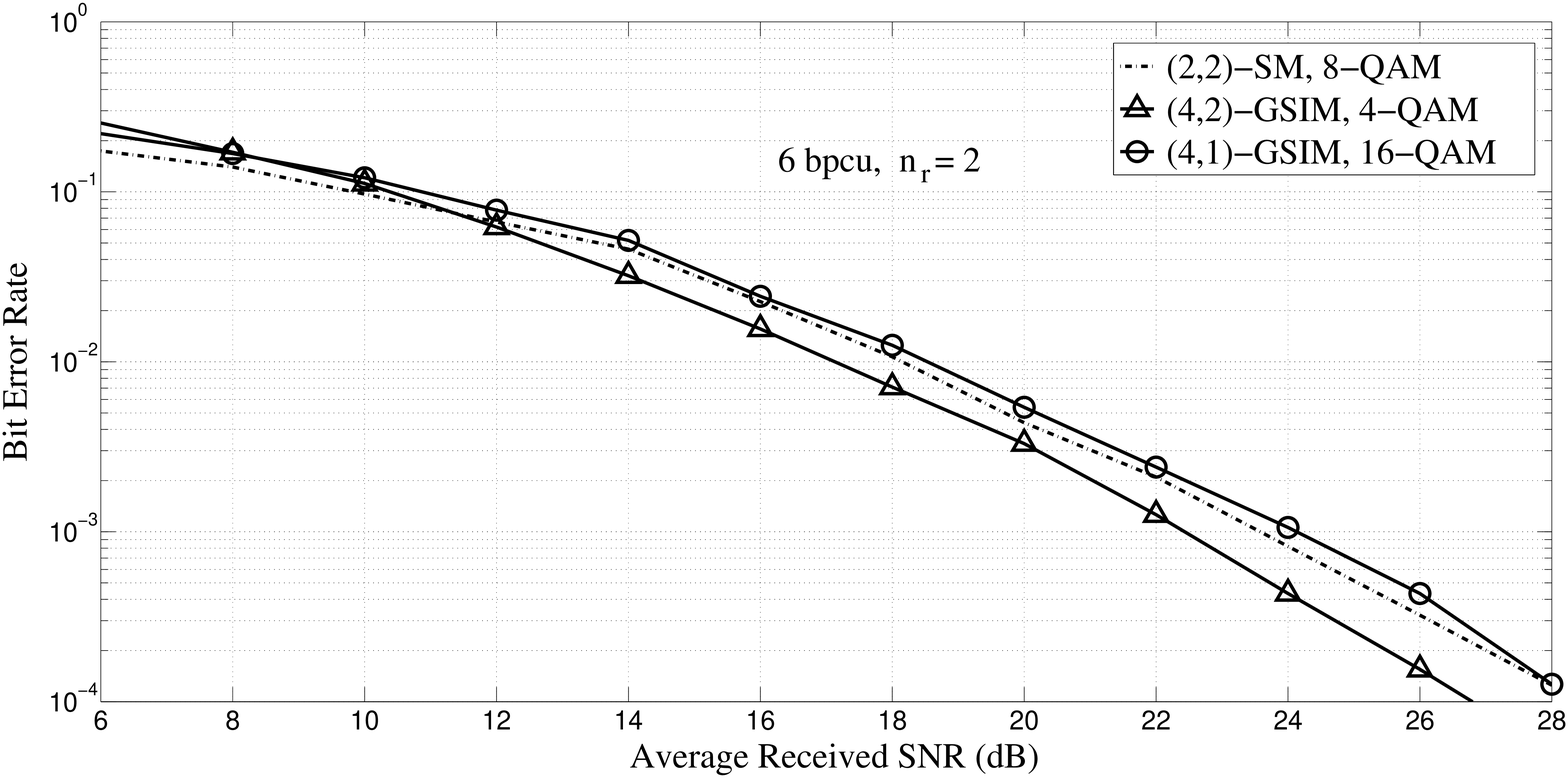}
\vspace{-2mm}
\caption{BER comparison between $(4,2)$-GSIM, $(4,1)$-GSIM, and
$(2,2)$-SM systems with 6 bpcu, $n_r=2$, and brute-force ML detection. }
\label{fig6_7}
\end{figure}

\begin{figure}
\centering
\includegraphics[height=2.5in,width=3.0in]{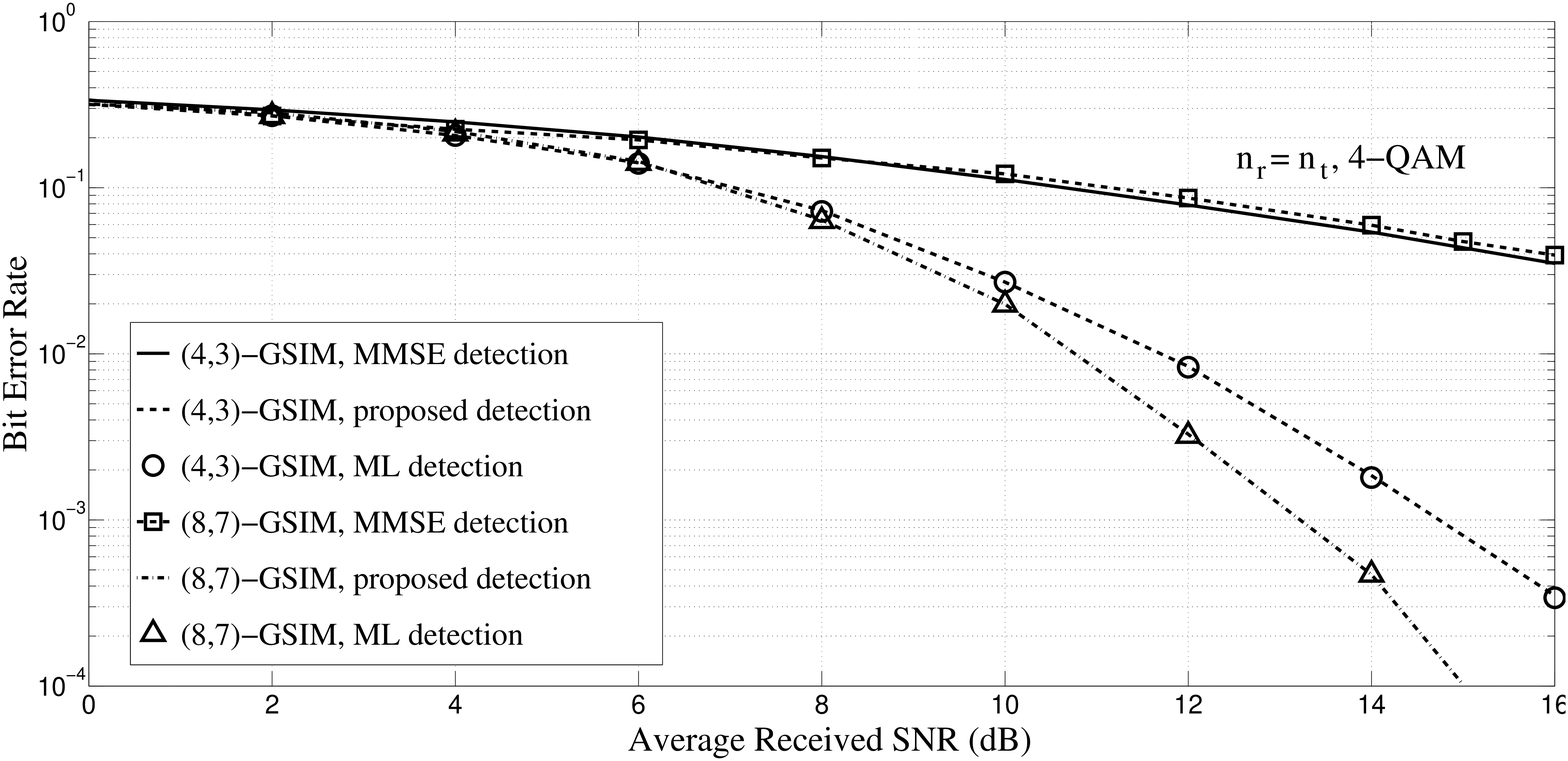}
\vspace{-2mm}
\caption{BER comparison between MMSE detection, proposed detection, and
brute-force ML detection in (4,3)-GSIM and (8,7)-GSIM systems with
$n_r=n_t$, and 4-QAM. }
\vspace{-4mm}
\label{fig6_8}
\end{figure}

\begin{figure}
\centering
\includegraphics[height=2.5in,width=3.0in]{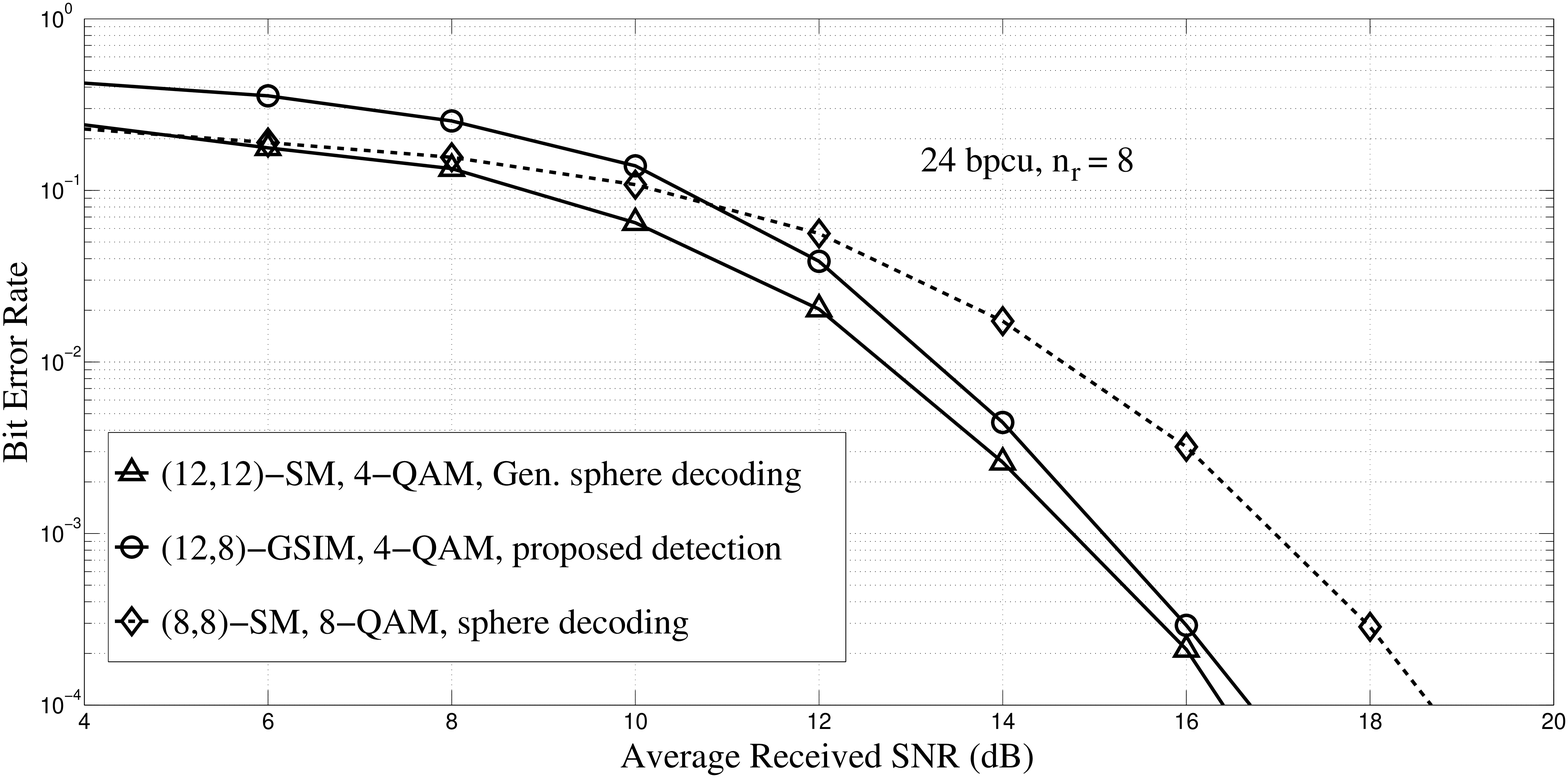}
\vspace{-2mm}
\caption{BER comparison among three systems achieving 24 bpcu: $i)$
(8,8)-SM system with 8-QAM, $ii)$ (12,8)-GSIM system with 4-QAM,
and $iii)$ (12,12)-SM system with 4-QAM, $n_r=8$.}
\label{fig6_9}
\end{figure}

\begin{figure}
\centering
\includegraphics[height=2.5in,width=3.0in]{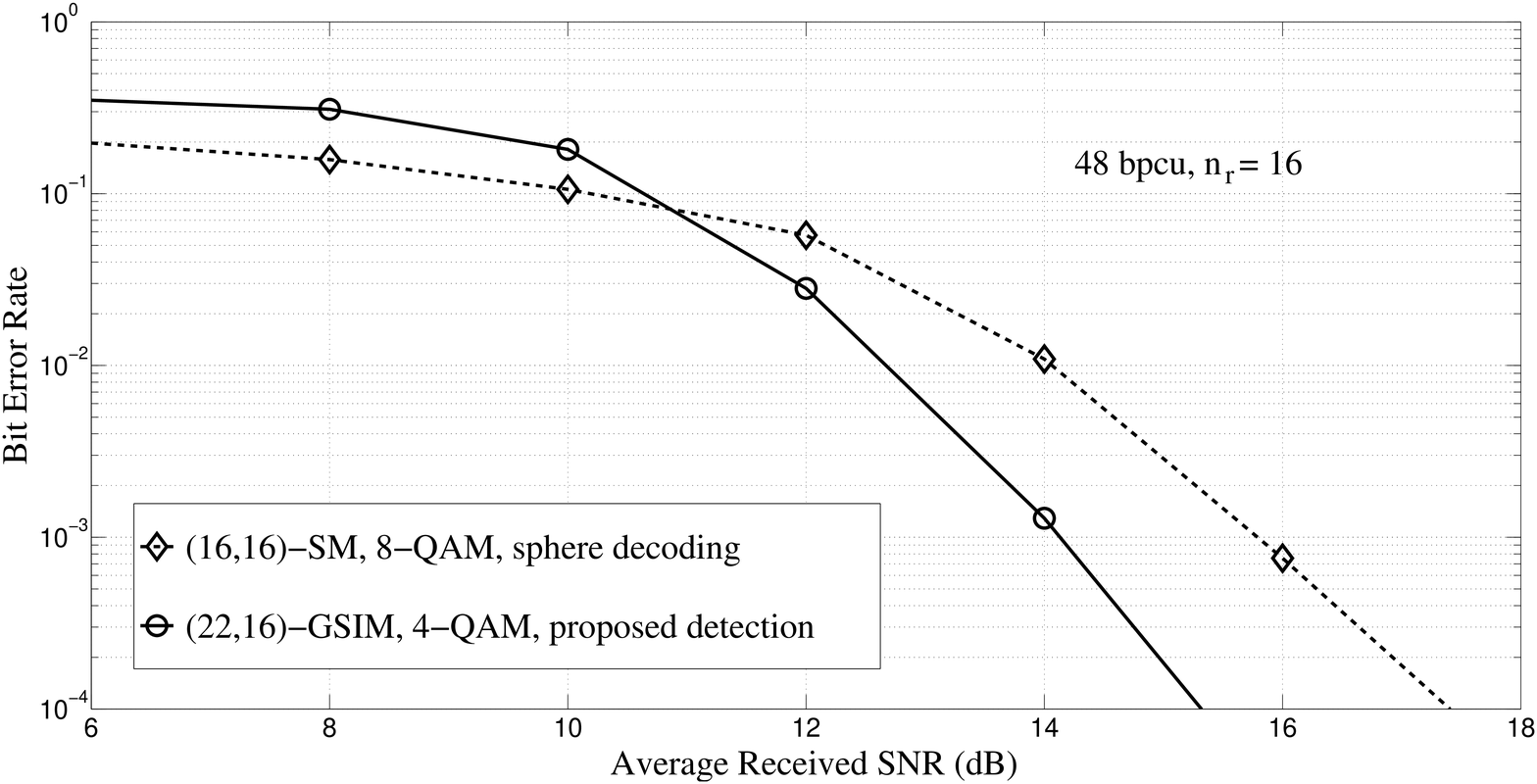}
\vspace{-2mm}
\caption{BER comparison between GSIM and SM systems using same RF
resources for $n_{rf}=n_{rf}^{opt}$, $n_r=n_{rf}$ to achieve 48 bpcu.}
\vspace{-4mm}
\label{fig6_10}
\end{figure}

In Fig. \ref{fig6_7}, we show the BER comparison between $i)$ $(4,2)$-GSIM
with 4-QAM, $ii)$ $(4,1)$-GSIM with 16-QAM, and $iii)$ $(2,2)$-SM with
8-QAM, using $n_r=2$. Note that in all the three systems, the modulation
alphabets have been chosen such that the rate is the same 6 bpcu. Since the
systems are small, brute-force ML detection is used. It can be seen that
(4,2)-GSIM system performs better than (2,2)-SM system. That is, for the
same rate of 6 bpcu and $n_{rf}=2$, GSIM achieves better performance than
spatial multiplexing by about 1 dB better performance at 0.01 uncoded BER.
As we will see in Figs. \ref{fig6_9} and \ref{fig6_10}, this
improvement increases to about 1.5 to 2 dB for 24 bpcu and 48 bpcu systems.
It is noted that GSIM needs extra transmit antennas than spatial multiplexing
to achieve this improvement. But the additional resources used in GSIM are
not the transmit RF chains (which are expensive), but only the transmit
antenna elements (which are not expensive). It can also be seen that
even (4,1)-GSIM performs close to within 0.5 dB of (2,2)-SM
performance in medium to high SNRs. This shows that GSIM can save RF
transmit chains without losing much performance compared to spatial
multiplexing.

Fig. \ref{fig6_8} shows the BER performance of different detection schemes
for GSIM. (4,3)-GSIM and (8,7)-GSIM with $n_r=n_t$ and 4-QAM are considered.
Note that the choice of $n_{rf}$ in both systems corresponds to $n_{rf}^{opt}$.
Three detectors, namely, MMSE detector, proposed detector, and brute-force
ML detector are considered. It can be seen that MMSE detector yields very
poor performance, but the proposed detector yields a performance which
almost matches the ML detector performance. The proposed detector achieves
this almost ML performance in just cubic complexity in $n_t$, whereas ML
detection has exponential complexity in $n_t$.

In Fig. \ref{fig6_9}, we compare the performance of three systems, each
achieving 24 bpcu: $i)$ (8,8)-SM with 8-QAM and ML detection using sphere
decoder (SD), $ii)$ (12,8)-GSIM with 4-QAM and proposed detection, and
$iii)$ (12,12)-SM system with 4-QAM using generalized sphere decoder
(GSD)\footnote{Since $n_r=8$, the (12,12)-SM system is an
underdetermined system. Therefore, we have used the GSD in \cite{gsd}
which achieves ML detection in such underdetermined systems.
GSD for spatial modulation has been reported in \cite{gsd2}.}.
All the three systems use $n_r=8$. Fig. \ref{fig6_9}
shows that the (12,8)-GSIM with proposed detection outperforms (8,8)-SM
with SD employing same RF resources by about 2 dB in high SNR regime
by using four extra transmit antennas.
The performance of (12,8)-GSIM with proposed detection is very close
to that of (12,12)-SM system with GSD which uses more RF resources to
achieve the same rate. Also, the proposed detector has a much lower
complexity than GSD which has exponential complexity in $n_t$.

Fig. \ref{fig6_10} shows the BER comparison between GSIM and SM
using same RF resources for $n_{rf}=n_{rf}^{opt}$, $n_r=n_{rf}$
to achieve 48 bpcu. GSIM uses $n_t=22$ and 4-QAM, whereas  (16,16)-SM
scheme uses 8-QAM modulation alphabet to match the rate. For GSIM,
the proposed detection is used. For SM, sphere decoding is used.
It can be seen that, (22,16)-GSIM scheme outperforms (16,16)-SM scheme
using same RF resources by about 2 dB in the medium to high SNR regime
by using six extra transmit antennas.
Also, the proposed detection has a much lower complexity than SD.

In Figs. \ref{fig6_9} and \ref{fig6_10}, we also observe that at
low SNRs the SM schemes have better BER performance compared to the
corresponding GSIM schemes. This can be explained as follows. First,
it can be observed that, to achieve the same rate, GSIM needs
smaller-sized constellation compared to SM. Hence, GSIM will have
a larger minimum distance among the constellation points than that
in SM. Second, unlike in SM where there are no antenna index bits, the
following two types of error events are observed in GSIM: $i$) the
antenna activity pattern itself is decoded wrongly, and thus both the
antenna index bits and modulation symbol bits are incorrectly decoded,
and $ii$) the antenna activity pattern is decoded correctly, but the
modulation symbol bits are wrongly decoded. At medium to high SNRs, the
error event of the second type is more likely to occur and therefore
this type of error events dominates the resulting performance. Coupled
with this, a larger minimum distance among constellation points in GSIM
than that in the corresponding SM makes GSIM to outperform SM in medium
to high SNRs. But at low SNRs, the error event of the first type is more
likely to occur and this error event type dominates the resulting
performance. Since there are no antenna index bits in SM, error events
of the first type do not occur in SM, leading to better performance for
SM in the low SNR regime.

\section{Generalized Space-Frequency Index Modulation}
\label{sec_gsfm}
In this section, we propose a generalized space-frequency index modulation
(GSFIM) scheme which encodes bits through indexing in both spatial as well
as frequency domains. GSFIM can be viewed as a generalization
of the GSIM scheme presented in the previous section by exploiting indexing
in the frequency domain as well.
In the proposed GSFIM scheme, information bits are
mapped through antenna indexing in the spatial domain, frequency indexing
in the frequency domain, and $M$-ary modulation. After mapping, the signal
is modulated using OFDM and is transmitted through the selected antennas. 
We obtain the rate equation for the proposed GSFIM system and study its
achievable rate, rate variation as a function of the parameters involved,
and the rate gain compared to conventional MIMO-OFDM.

\subsection{System model}
\label{chp4_sec2}
\begin{figure}
\centering
\includegraphics[height=2.25in, width=3.4in]{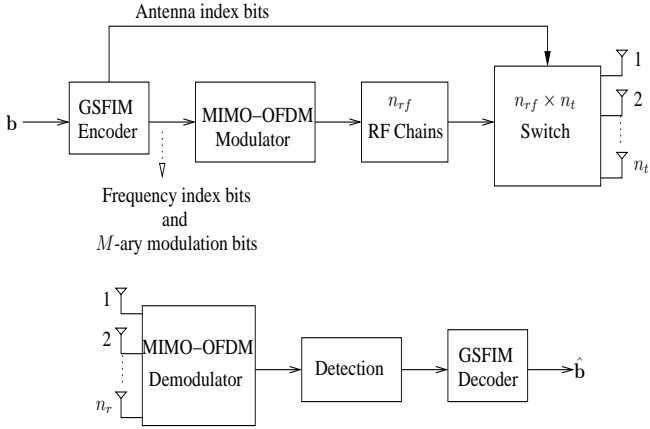}
\vspace{-0mm}
\caption{Block diagram of GSFIM transmitter and receiver.}
\label{fig_gsfm1}
\vspace{-4mm}
\end{figure}

The proposed GSFIM system uses $n_t$ transmit antennas, $n_{rf}$ transmit
RF chains, $1\leq n_{rf}\leq n_t$, $N$ subcarriers, and $n_r$ receive
antennas. The channel between each transmit and receive antenna pair is
assumed to be frequency-selective fading with $L$ multipaths.
The block diagrams of the GSFIM transmitter and receiver are
shown in Fig. \ref{fig_gsfm1}. At any given time, only $n_{rf}$ transmit
antennas are active and the remaining $n_t-n_{rf}$ antennas remain silent.
The GSFIM encoder takes $\lfloor \log_2{n_t\choose n_{rf}}\rfloor$ bits and
maps to $n_{rf}$ out of $n_t$ transmit antennas (antenna index bits).
It also takes additional bits to index subcarriers (frequency index bits)
and bits for $M$-ary modulation symbols on subcarriers. The frequency and
antenna indexing mechanisms are detailed below.

\subsubsection{Frequency indexing}
\begin{figure}
\hspace{-2.4mm}
\includegraphics[height=1.0in, width=3.5in]{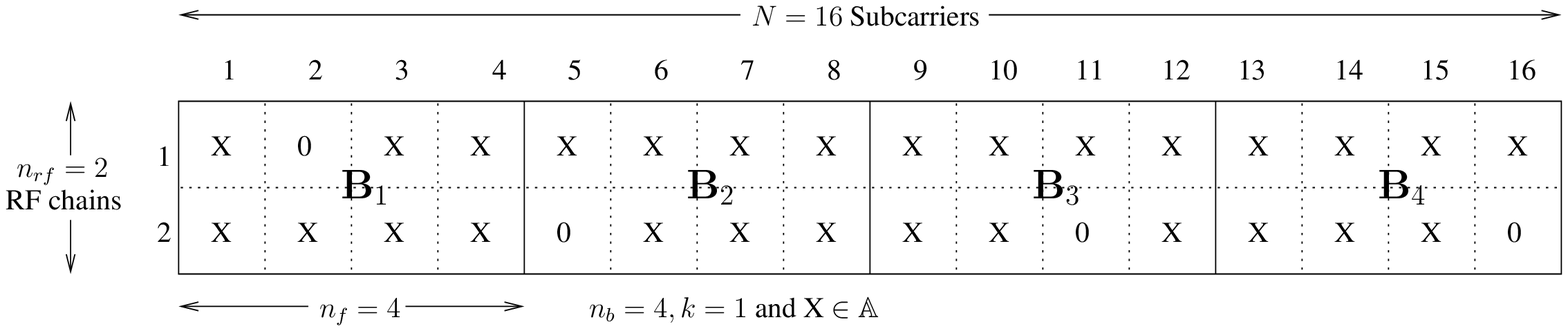}
\vspace{-0mm}
\caption{Frequency indexing in GSFIM.}
\vspace{-4mm}
\label{fig_gsfm2}
\end{figure}

Consider a matrix ${\bf B}$ of size $n_{rf} \times N$ whose entries belong
to ${\mathbb A}_0$, where ${\mathbb A}_0 = {\mathbb A} \cup 0$ with
${\mathbb A}$ denoting an $M$-ary modulation alphabet. The frequency
index bits and $M$-ary modulation bits are embedded in ${\bf B}$ as
follows. The matrix ${\mathbf B}$ is divided into $n_b$ sub-matrices
${\mathbf B}_1, {\mathbf B}_2, \cdots {\mathbf B}_{n_b},$ each of size
$n_{rf} \times n_f$, where $n_f=\frac{N}{n_b}$ is the number subcarriers per
sub-matrix (see Fig. \ref{fig_gsfm2}).  Let $k$, $1\leq k \leq n_{rf}n_f$
denote the number of non-zero elements in each sub-matrix, where each of
the non-zero elements belong to ${\mathbb A}$. This $k$ is a design
parameter. Then, for each sub-matrix, there are $l_f={n_{rf}n_f \choose k}$
possible `frequency activation patterns'. A frequency activation pattern
for a given sub-matrix refers to a possible combination of zero and
non-zero entries in that sub-matrix. Note that not all $l_f$ activation
patterns are needed for frequency indexing. Any $2^{K_f}$ patterns out of
them, where $k_f=\left \lfloor \log_2{n_{rf}n_f \choose k} \right\rfloor$,
are adequate. Take any $2^{K_f}$ patterns out of $l_f$ patterns and form
a set called the `frequency activation pattern set', denoted by
${\mathbb S}_f$. The frequency activation pattern for a given sub-matrix
is then formed by choosing one among the patterns in the set ${\mathbb S}_f$
using $k_f$ bits. These $k_f$ bits are the frequency index bits for that
sub-matrix. So, there are a total of $n_bk_f$ frequency index bits in the
entire matrix ${\bf B}$. In addition to these frequency index bits,
$kn_b\log_2M$ bits are carried as $M$-ary modulation bits in the non-zero
entries of ${\bf B}$.

{\em Example:}
Let us illustrate this using the following example. Let $n_{rf}=2$, $N=16$,
$n_b=4$, and $k=7$. Then, $n_f=\frac{16}{4}=4$, $l_f={8 \choose 7}=8$,
$k_f = \left \lfloor \log_2 8 \right \rfloor = 3$, and $2^{k_f}=8$.
In this example, $l_f=2^{K_f}=8$, i.e., all the 8 possible patterns
are in the frequency activation pattern set, given by

\vspace{-2mm}
{\small
\begin{eqnarray*}
{\mathbb S}_f & \hspace{-2mm} = & \hspace{-2mm} \left\{
\begin{bmatrix}
0& 1& 1& 1 \\
1& 1& 1& 1
\end{bmatrix},
\begin{bmatrix}
1& 0& 1& 1 \\
1& 1& 1& 1
\end{bmatrix},
\begin{bmatrix}
1& 1& 0& 1 \\
1& 1& 1& 1
\end{bmatrix}, \right. \nonumber \\
& & \hspace{1mm}
\begin{bmatrix}
1& 1& 1& 0 \\
1& 1& 1& 1
\end{bmatrix}, 
\begin{bmatrix}
1& 1& 1& 1 \\
0& 1& 1& 1
\end{bmatrix},
\begin{bmatrix}
1& 1& 1& 1 \\
1& 0& 1& 1
\end{bmatrix}, \nonumber \\
& & \hspace{1mm}\left.
\begin{bmatrix}
1& 1& 1& 1 \\
1& 1& 0& 1
\end{bmatrix},
\begin{bmatrix}
1& 1& 1& 1 \\
1& 1& 1& 0
\end{bmatrix}
\right\}.
\end{eqnarray*}
}

\vspace{-2mm}
\hspace{-5mm}
Suppose ${\mathbb A}$ is 4-QAM. Let
$[0 0 1 0 1 0 0 1 1 1 1 0 0 0 1 1 0]$ denote the information bit sequence
for sub-matrix ${\bf B}_1$. The GSFIM encoder translates these bits to the
sub-matrix $\mathbf {B}_1$ as follows: the first 3 bits are used to choose
the frequency activity pattern (i.e., 001 chooses the activation pattern
{\scriptsize
$
\begin{bmatrix}
1& 0& 1& 1 \\
1& 1& 1& 1
\end{bmatrix}
$
}
in the set ${\mathbb S}_f$ above), and the next 14 bits are mapped to
seven 4-QAM symbols so that one 4-QAM symbol gets mapped to one active
subcarrier. The sub-matrix ${\bf B}_1$ then becomes
\[
{\bf B}_1 \ = \
\begin{bmatrix}
-1-{\bf j} & 0& -1+{\bf j} & 1-{\bf j} \\
1-{\bf j} & -1+{\bf j} & -1-{\bf j} & 1+{\bf j}
\end{bmatrix},
\]
where {\small ${\bf j}=\sqrt{-1}$}. Likewise, the sub-matrices
$\mathbf{B}_i$, $i=2,3,4$ are formed. The full matrix ${\mathbf B}$
of size $n_{rf}\times N$ is then formed as
\[
{\mathbf B} \ = \ [\mathbf{B}_1 \ \mathbf{B}_2 \ \mathbf{B}_3 \ \mathbf{B}_4].
\]
Each row of the matrix ${\mathbf B}$ is of dimension $1\times N$. There are
$n_{rf}$ rows. Each $N$-length row vector in ${\mathbf B}$ is fed to the
IFFT block in the OFDM modulator to generate an $N$-length OFDM symbol.
A total of $n_{rf}$ such OFDM symbols, one for each row in ${\mathbf B}$,
are generated. These $n_{rf}$ OFDM symbols are then transmitted through
$n_{rf}$ active transmit antennas in parallel.
The choice of these $n_{rf}$ active transmit antennas among the $n_t$
available antennas is made through antenna indexing as described below.

\subsubsection{Antenna indexing}
The selection of $n_{rf}$ out of $n_t$ antennas for transmission is made
based on antenna index bits. The antenna index bits choose an `antenna
activation pattern', which tells which $n_{rf}$ antennas out of $n_t$
antennas are used for transmission. There are $l_a={n_t\choose n_{rf}}$
antenna activation patterns possible, and
$k_a = \big \lfloor \log_2{n_t \choose n_{rf}} \big \rfloor$ bits are
used to choose one among them. These $k_a$ bits are the antenna
index bits. Note that not all $l_a$ activation patterns are needed,
and any $2^{k_a}$ patterns out of them are adequate. Take any $2^{k_a}$
patterns out of $l_a$ patterns and form a set called the `antenna
activation pattern set', denoted by ${\mathbb S}_a$.

{\em Example:}
Let us illustrate this using the following example. Let $n_t=3$, $n_{rf}=2$.
Then, $l_a={3\choose 2}=3$,
$k_a=\left\lfloor\log_2{3\choose 2}\right\rfloor=\lfloor\log_2 3\rfloor=1$,
and $2^{k_a}=2$. The possible antenna activation patterns are given
by $\big\{[ 1, 1, 0]^T, [1, 0, 1 ]^T, [ 0, 1, 1 ]^T \big\}$. The set
${\mathbb S}_a$ is formed by selecting any two patterns out of the
above three patterns. For example, ${\mathbb S}_a$ can be
\[
{\mathbb S}_a = \big\{[ 1, 1, 0]^T, [1, 0, 1 ]^T\big\}.
\]
An $n_{rf}\times n_t$ switch connects the transmit RF chains to the
transmit antennas. The chosen $n_{rf}$ out of $n_t$ transmit antennas
transmit the MIMO-OFDM symbol constructed using the frequency index
bits and $M$-ary modulation bits. The active transmit antennas can
change from one MIMO-OFDM symbol to the other.

\subsection{Achievable rate, rate variation, and rate gain}
\label{chp4_sec3}
In GSFIM, the information bits are encoded using $i)$ frequency indexing
over each sub-matrix ${\bf B}_i$, $i=1,2,\cdots,n_b$, $ii)$ $M$-ary
modulation symbols in each sub-matrix, and $iii)$ antenna indexing.
The number of frequency indexing bits per sub-matrix is
$\left\lfloor \log_2{n_{rf}n_f \choose k}\right\rfloor$. The
number of $M$-ary modulation bits in each sub-matrix is $k\log_2M$.
The number of antenna indexing bits is
$\left\lfloor \log_2{n_t \choose n_{rf}}\right\rfloor$. Combining these
three parts, the achievable rate in GSFIM with $n_t$ transmit antennas,
$n_{rf}$ transmit RF chains, $N$ subcarriers, $n_b$ sub-matrices, and
$M$-ary modulation is given by
\begin{eqnarray}
R_{\mbox{{\scriptsize gsfim}}}&=&
\underbrace{\left(\frac{\left\lfloor \log_2{n_t \choose n_{rf}}\right\rfloor}{N+L-1}\right)}_{R_{\text A}}
+\underbrace{\left(\frac{\left\lfloor \log_2{n_{rf}n_f \choose k}\right\rfloor n_b}{N+L-1}\right)}_{R_{\text F}} \nonumber \\
& \hspace{-2mm} & \hspace{-2mm} + \underbrace{\left(\frac{kn_b\log_2 M}{N+L-1}\right)}_{R_{\text Q}}
\quad \mbox{bpcu}.
\label{gsfm_rate1}
\end{eqnarray}

Note that in a conventional MIMO-OFDM system, there is no contribution
to the rate by antenna or frequency indexing, and the achieved rate is
only through $M$-ary modulation symbols. Also, in MIMO-OFDM, $M$-ary
modulation symbols are mounted on all $N$ subcarriers on each of the
$n_{rf}$ active transmit antennas. Therefore, the achieved rate in
MIMO-OFDM (with no antenna and frequency indexing) for the same
parameters as in GSFIM is given by
\begin{eqnarray}
R_{\mbox{{\scriptsize mimo-ofdm}}}&\hspace{-2mm} = &\hspace{-2mm}
\left(\frac{1}{N+L-1}\right) n_{rf}N\log_2M \quad \mbox{bpcu}.
\label{ofdm_rate}
\end{eqnarray}

\begin{figure}
\subfigure[$M=2$]{\includegraphics[height=2.5in, width=3.5in]{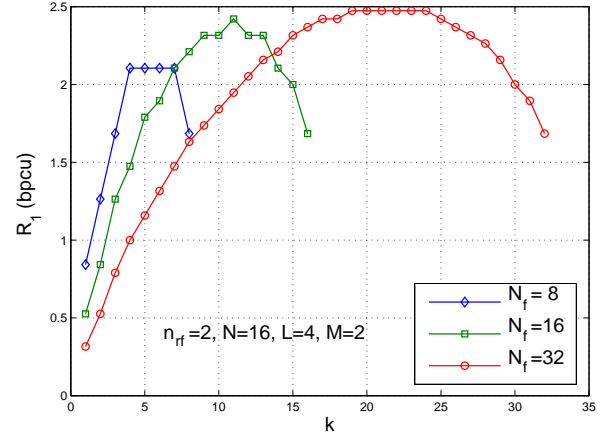}}
\subfigure[$M=4$]{\includegraphics[height=2.5in, width=3.5in]{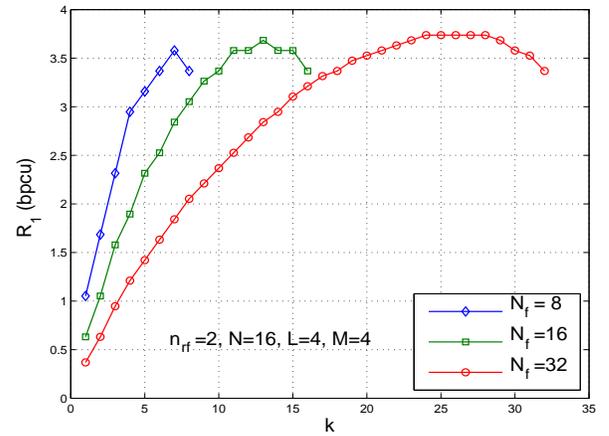}}
\caption{Rate $R_1=R_{\text F}+R_{\text Q}$ as a function of $k$
for different values of $N_f=n_{rf}n_f$. }
\label{bps_Nf}
\end{figure}

From (\ref{gsfm_rate1}) and (\ref{ofdm_rate}), we can make the
following observations:
\begin{itemize}
\item conventional MIMO-OFDM becomes a special case of GSFIM for
$n_{rf}=n_t$, $n_f=N$ (i.e., $n_b=1$).

\item GSIM presented in Section \ref{gsm_1} becomes a special case of
GSFIM for $N=n_f=n_b=1$, $k=n_{rf}$.

\item for $n_{rf} < n_t$, $R_{\text A} > 0$, which is the additional
rate contributed by antenna indexing. In this case,
$R_{\mbox{\scriptsize gsfim}}$ in (\ref{gsfm_rate1}) can be more or less
compared to $R_{\mbox{{\scriptsize mimo-ofdm}}}$  depending on the
choice of parameters. For example, the parameter $k$ can take values
in the range 1 to $n_{rf}n_f$. An instance where
$R_{\mbox{\scriptsize gsfim}}$ is more than
$R_{\mbox{{\scriptsize mimo-ofdm}}}$ happens when
$k=n_{rf}n_f$, in which case $R_{\text F}=0$ and
$R_{\text Q}=R_{\mbox{{\scriptsize mimo-ofdm}}}$. Therefore,
$R_{\text A}$ is the excess rate (rate gain) in GSFIM compared to
MIMO-OFDM. Likewise, an instance where $R_{\mbox{\scriptsize gsfim}}$
is less than $R_{\mbox{{\scriptsize mimo-ofdm}}}$ happens when $k=1$,
in which case $R_{\mbox{\scriptsize gsfim}}$ becomes
$\frac{n_b\log_2(n_{rf}n_fM)}{N+L-1}$ which is less than
$R_{\mbox{{\scriptsize mimo-ofdm}}}$ given by
$\frac{n_bn_{rf}n_f \log_2(M)}{N+L-1}$.

\item the sum of rates $R_{\text F}$ and $R_{\text Q}$ in (\ref{gsfm_rate1})
as a function of $k$ reaches its maximum for a value of $k$ in the range
$\lfloor \frac{n_{rf}n_f}{2}\rfloor$ and $n_{rf}n_f$, and so does the
total rate $R_{\mbox{{\scriptsize gsfim}}}$. The maximum
$R_{\mbox{{\scriptsize gsfim}}}$ will be more than or equal to
$R_{\mbox{{\scriptsize mimo-ofdm}}}$.

\end{itemize}
We now illustrate the above observations through numerical results.
Define $R_1 \Define R_{\text F}+R_{\text Q}$ and $N_f \Define n_{rf}n_f$.
In Fig.  \ref{bps_Nf}, we plot $R_1$ as a function of $k$, for different
values of $N_f=8,16,32$, $L=4$, and $M=2,4$. We observe that $R_1$ reaches
its maximum value for $k$ between $\lfloor \frac{N_f}{2} \rfloor$ and $N_f$.
Also, the maximum $R_1$ increases as $N_f$ increases because the $R_{\text F}$
term in (\ref{gsfm_rate1}) increases with $N_f$.

\begin{figure}
\subfigure[$M=2$]{\includegraphics[height=2.5in, width=3.5in]{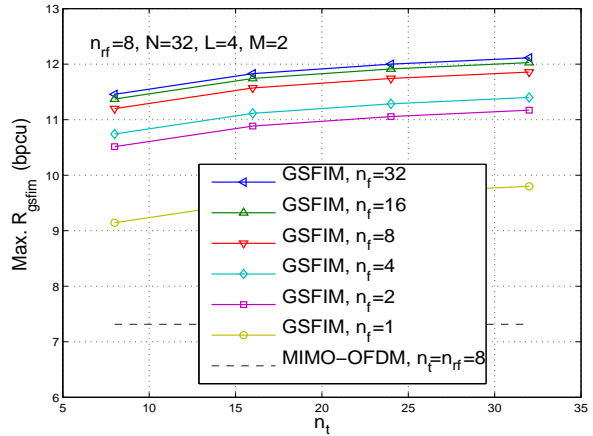}}
\subfigure[$M=4$]{\includegraphics[height=2.5in, width=3.5in]{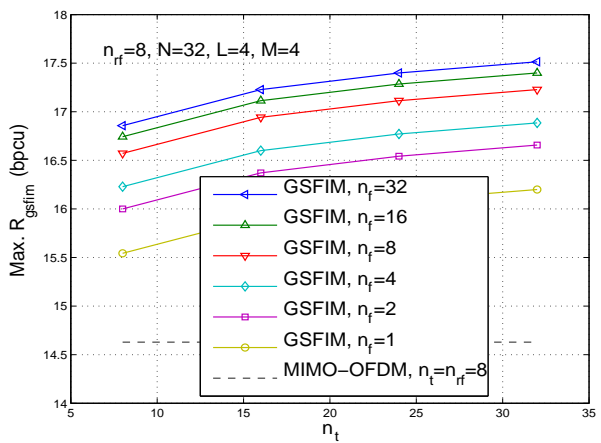}}
\caption{Maximum $R_{{\scriptsize \mbox{gsfim}}}$ as
a function of $n_t$, for $n_{rf}=8$ and different values of $n_{f}$.}
\label{bps_nt}
\end{figure}

\begin{figure}
\subfigure[$M=2$]{\includegraphics[height=2.5in, width=3.5in]{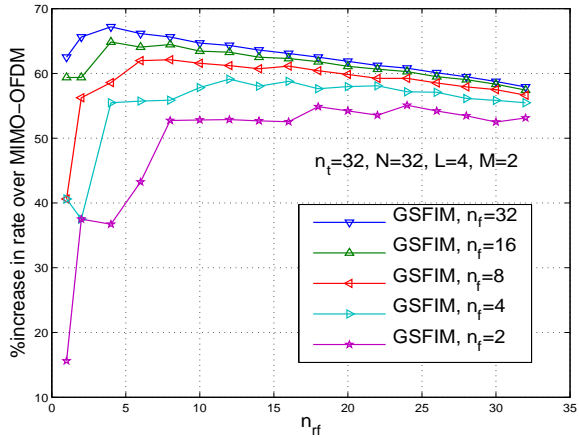}}
\subfigure[$M=4$]{\includegraphics[height=2.5in, width=3.5in]{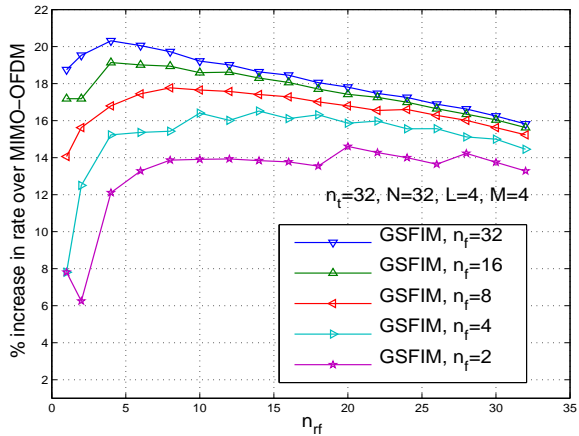}}
\caption{Percentage rate gain in GSFIM compared to MIMO-OFDM as a
function of $n_{rf}$ and $n_f$. }
\label{bps_per}
\end{figure}

In Fig. \ref{bps_nt}, we plot the maximum $R_{{\scriptsize \mbox{gsfim}}}$
as a function of $n_t$ for $n_{rf}=8$, $N=32$, $L=4$, and $n_f=1,2,4,8,16,32$.
$R_{{\scriptsize \mbox{mimo-ofdm}}}$ is also plotted for comparison. We
observe that for a given $n_f$, the maximum $R_{{\scriptsize \mbox{gsfim}}}$
increases with $n_t$ because of the increase in antenna index bits carried.
For a given $n_t$ and $n_{rf}$, the maximum $R_{{\scriptsize \mbox{gsfim}}}$
increases with increase in $n_f$ because of increase in $N_f$ and the
associated increase in $R_{\text F}$.
From this figure, we can see that GSFIM can achieve a rate gain of up to
65\% for $M=2$ and up to 19\% for $M=4$, compared to MIMO-OFDM.
In Fig. \ref{bps_per}, we have
plotted the percentage rate gain in GSFIM compared to MIMO-OFDM (i.e.,
difference between maximum $R_{\mbox{{\scriptsize gsfim}}}$ and
$R_{\mbox{{\scriptsize mimo-ofdm}}}$ in percentage), as a function of
$n_{rf}$ for $n_t=32$, $N=32$, $L=4$, and $n_f=2,4,8,16,32$. As can be
observed in Fig. \ref{bps_per}, GSFIM can achieve rate gains up to 65\%
for $M=2$ and 20\% for $M=4$, compared to MIMO-OFDM.

\begin{figure}
\subfigure[$M=2$]{\includegraphics[height=2.5in, width=3.5in]{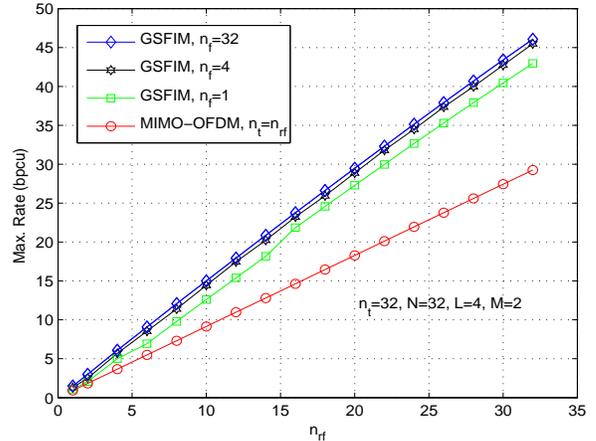}}
\subfigure[$M=4$]{\includegraphics[height=2.5in, width=3.5in]{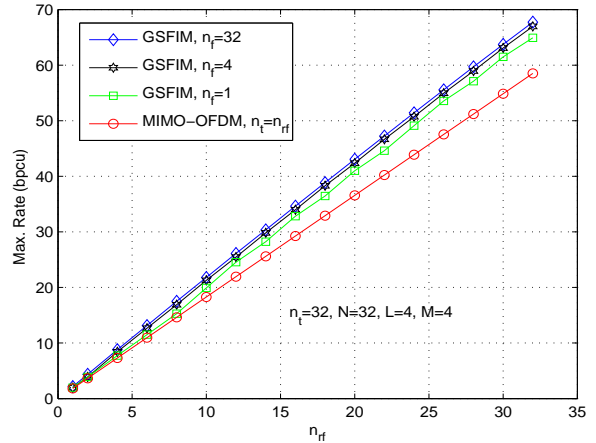}}
\caption{Maximum $R_{\mbox{{\scriptsize gsfim}}}$ as a function of
$n_{rf}$, for $n_t=N=32$, and $n_f=1,4,32$.}
\label{bps_nrf}
\end{figure}

\begin{figure}
\subfigure[$M=2$]{\includegraphics[height=2.5in, width=3.5in]{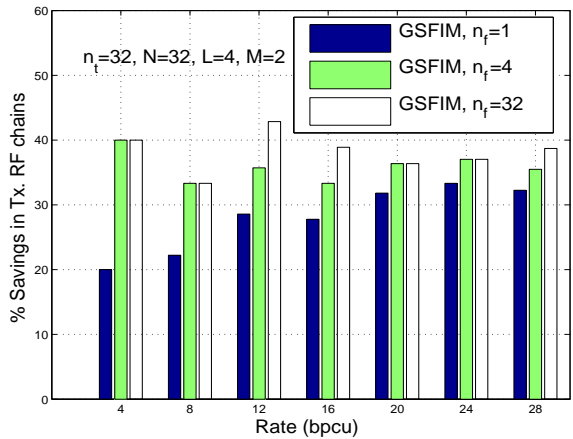}}
\subfigure[$M=4$]{\includegraphics[height=2.5in, width=3.5in]{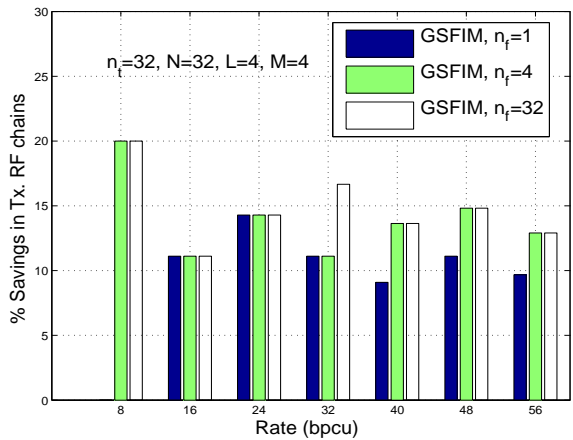}}
\caption{Percentage savings in number of transmit RF chains in GSFIM
compared to MIMO-OFDM, for $n_t=N=32$, $n_f=1,4,32$.  }
\label{bps_rf}
\vspace{-4mm}
\end{figure}

In Fig. \ref{bps_nrf}, we plot the maximum $R_{\mbox{{\scriptsize gsfim}}}$
as a function of $n_{rf}$ for a given $n_t=32$, $N=32$, $L=4$ and $n_f=1,32$.
We can observe that for a given $n_f$, the rate increases with $n_{rf}$
because of the increase in $R_{\text F}$. For a given $n_{rf}$, the
the maximum $R_{\mbox{{\scriptsize gsfim}}}$ increases with increase in
$n_f$. In Fig. \ref{bps_rf}, we have plotted bar graphs showing the
percentage savings in transmit RF chains in GSFIM compared to MIMO-OFDM
$n_t=N=32$, $L=4$, and $n_f=1,4,32$.  It can be observed that this savings
is high for small-sized modulation alphabets -- e.g., the savings
is up to 42\% for $M=2$ and 20\% for $M=4$.

\begin{figure}
\centering
\label{bps_nf}
\includegraphics[height=2.75in, width=3.5in]{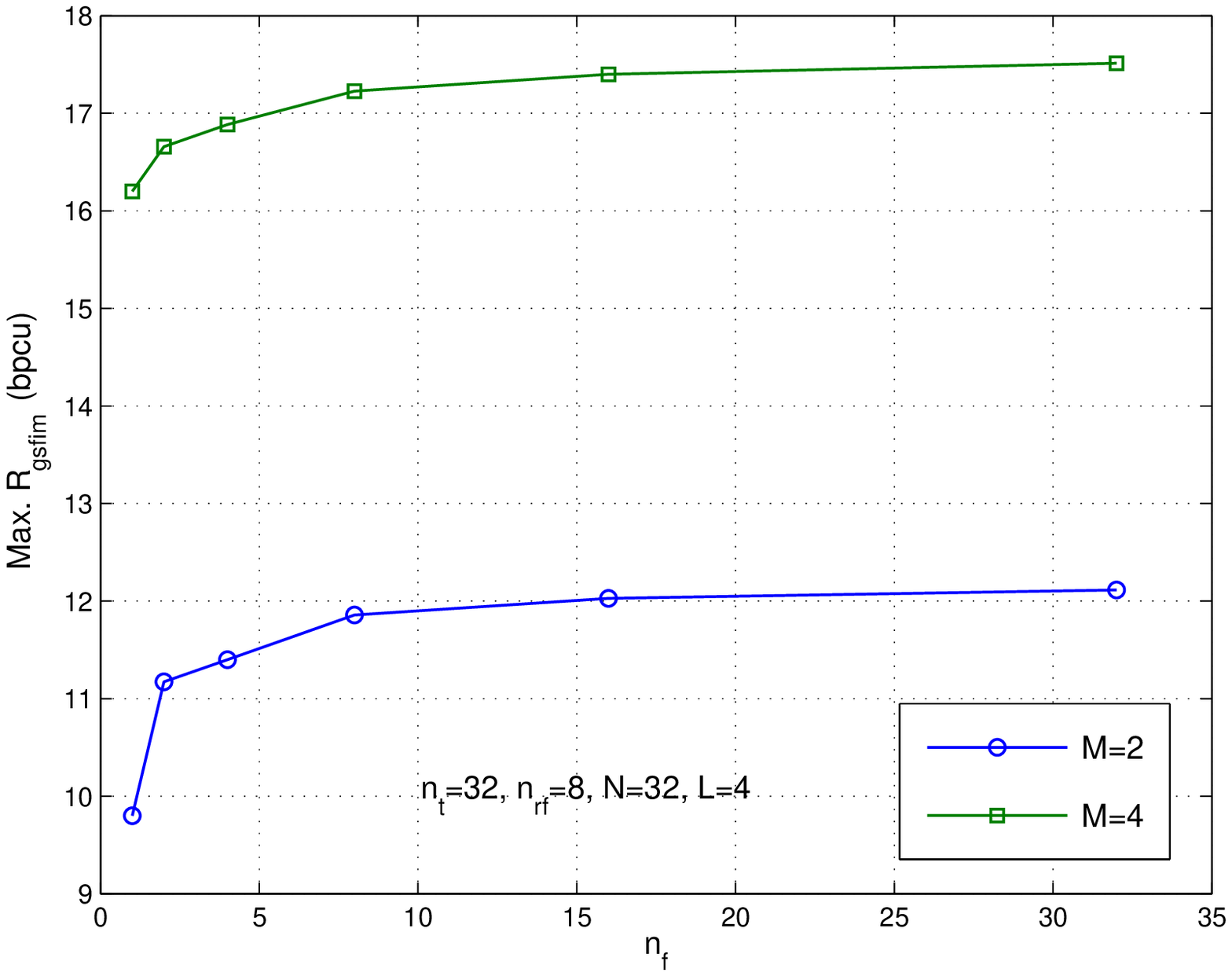}
\vspace{-6mm}
\caption{Maximum $R_{\mbox{{\scriptsize gsfim}}}$ as a function of
$n_f$, for fixed $n_t, n_{rf}$.}
\label{fig_gsm1a}
\vspace{-4mm}
\end{figure}

In Fig. \ref{fig_gsm1a}, we plot the maximum $R_{\mbox{{\scriptsize gsfim}}}$
as a function of $n_f$ for $n_t=32$, $n_{rf}=8$, $L=4$, and $N=32$. We can
observe that the maximum $R_{\mbox{{\scriptsize gsfim}}}$ increases for up
to certain $n_f$ and thereafter it saturates. This is because the maximum
$R_1$ saturates to a value $\frac{N_fn_b \log_2(M+1)}{N+L-1}$ for large $N_f$.

\subsection{GSFIM signal detection and performance}
\label{sec_gsfim_det}
In this subsection, we consider GSFIM signal detection and performance.
Let ${\bf H}_n$ denote $n_r\times n_t$ channel matrix on subcarrier $n$.
Let $\mathbf {H}^{\bf a}_n$ denote the $n_r \times n_{rf}$ channel matrix
corresponding to  the chosen $n_{rf}$ antennas.
The superscript ${\bf a}$ in $\mathbf {H}^{\bf a}_n$ refers to the
antenna activation pattern that tells which $n_{rf}$ antennas are chosen.
Let us denote the
$n_r\times 1$-sized received vector on subcarrier $n$ as
${\mathbf {y}_n}$, which can be written as
\begin{eqnarray}
{\mathbf y_n}={\mathbf H}^{\bf a}_n{\mathbf z}_n+{\mathbf w}_n, \quad n=1,2,\cdots,N,
\label{eq1}
\end{eqnarray}
where ${\mathbf z_n}$ is the $n_{rf}\times1$-sized transmitted vector on
subcarrier $n$, and ${\mathbf w_n}$ is the $n_r\times 1$-sized additive
white Gaussian noise vector at the receiver,
$\mathbf{w}_n \sim \mathcal{CN}(0,\sigma^2\mathbf{I}_{n_r})$.
Consider the system model in (\ref{eq1}) for the $i$th sub-matrix, given by
\begin{eqnarray}
\mathbf{y}_l & = & \mathbf{H}^{\bf a}_l \mathbf{z}_l + \mathbf{w}_l, \quad
l= i_1,i_2,\cdots,i_j,\cdots,i_{n_f},
\label{eq3}
\end{eqnarray}
where $i_j=(i-1)n_f+j$. Write (\ref{eq3}) as
\begin{eqnarray}
\label{ml_det1}
\mathbf{y}^i & = & \mathbf{G}^{\bf a}_i \mathbf{z}^i+\mathbf{w}^i, \quad
i=1,2,\cdots,n_b,
\end{eqnarray}
where
\[
\mathbf{{y}}^{i}=\left[
\begin{array}{c}
\mathbf{y}_{i_1}\\
\mathbf{y}_{i_2} \\
\vdots\\
\mathbf{y}_{i_{n_f}}
\end{array} \right], \quad
\mathbf{z}^{i}=\left[
\begin{array}{c}
\mathbf{z}_{i_1}\\
\mathbf{z}_{i_2} \\
\vdots\\
\mathbf{z}_{i_{n_f}}
\end{array} \right],
\]
\[ 
\mathbf{G}^{\bf a}_i=
\left[
\begin{array}{cccc}
\mathbf{H}^{\bf a}_{i_1}  &  &  & 0 \\
& \mathbf{H}^{\bf a}_{i_2} \\
&  &  \ddots \\
0  &  &  &  \mathbf{H}^{\bf a}_{i_{n_f}} \\
\end{array}
\right].
\]\\
The ML metric for a given antenna activation
pattern ${\bf a}$ and vectors ${\bf z}^i, i=1,\cdots,n_b$ representing
the frequency activation pattern and $M$-ary modulation bits is
\begin{eqnarray}
\label{ml_metric}
d(\mathbf a,\mathbf{z}^1,\mathbf{z}^2,\cdots,\mathbf{z}^{n_b}) & = & \sum_{i=1}^{n_b} \| \mathbf{y}^i - \mathbf{G}_i^{\mathbf a} \mathbf{z}^i \|^2.
\end{eqnarray}
Let ${\mathbb U}$ denote the set of all possible $N_f$-length transmit
vectors corresponding to a sub-matrix. Then, ${\mathbb U}$ is given by
\begin{eqnarray}
\mathbb{U} & = & \{\mathbf{x} | \mathbf{x}\in {\mathbb{A}_0}^{N_f\times 1},\|\mathbf{x}\|_0=k,{\bf t}^\mathbf{x}\in \mathbb{S}_f\},
\label{Udef1}
\end{eqnarray}
where ${\bf t}^{\bf x}$ denotes the
frequency activity pattern corresponding to ${\bf x}$, where
$t^{\bf x}_j=1, \text{iff} \, \, {x}_j\neq 0 \hspace{2mm},
\forall j=1,2,\cdots,N_f.$ The antenna activation and frequency activation
pattern sets ($\mathbb{S}_a$, $\mathbb{S}_f$), and the antenna and
frequency index bit maps are known at both transmitter and receiver.
Therefore, from (\ref{ml_metric}) and (\ref{Udef1}),
the ML decision rule for GSFIM signal detection is given by
\begin{eqnarray}
\label{ml_sol}
(\hat{\mathbf{a}},\hat{\mathbf{z}}^1,\hat{\mathbf{z}}^2,\cdots,\hat{\mathbf{z}}^{n_b})& = & \argmin_{\mathbf{a} \in \mathbb{S}_a,\, \mathbf{z}^i \in \mathbb{U}, \forall i} \ d(\mathbf{a},\mathbf{z}^1,\mathbf{z}^2,\cdots,\mathbf{z}^{n_b}).
\end{eqnarray}
By inverse mapping, the antenna index bits are recovered from
$\hat{\mathbf {a}}$ and the frequency index bits are recovered from
$\hat{\mathbf{z}}^1,\hat{\mathbf{z}}^2,\cdots,\hat{\mathbf{z}}^{n_b}$.

\begin{figure}
\subfigure[$N=8$]{\includegraphics[height=2.75in, width=3.5in]{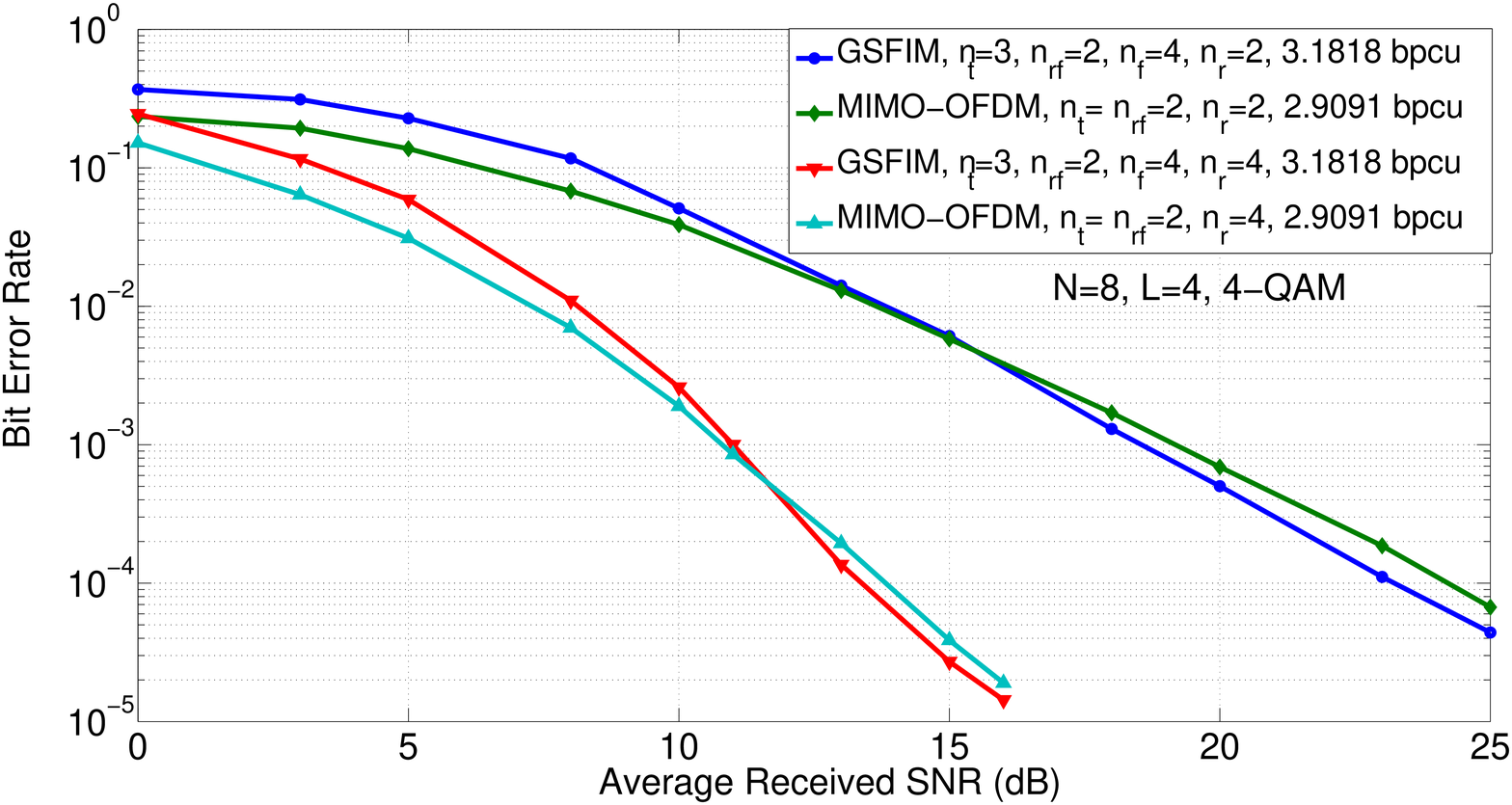}}
\subfigure[$N=16$]{\includegraphics[height=2.75in, width=3.5in]{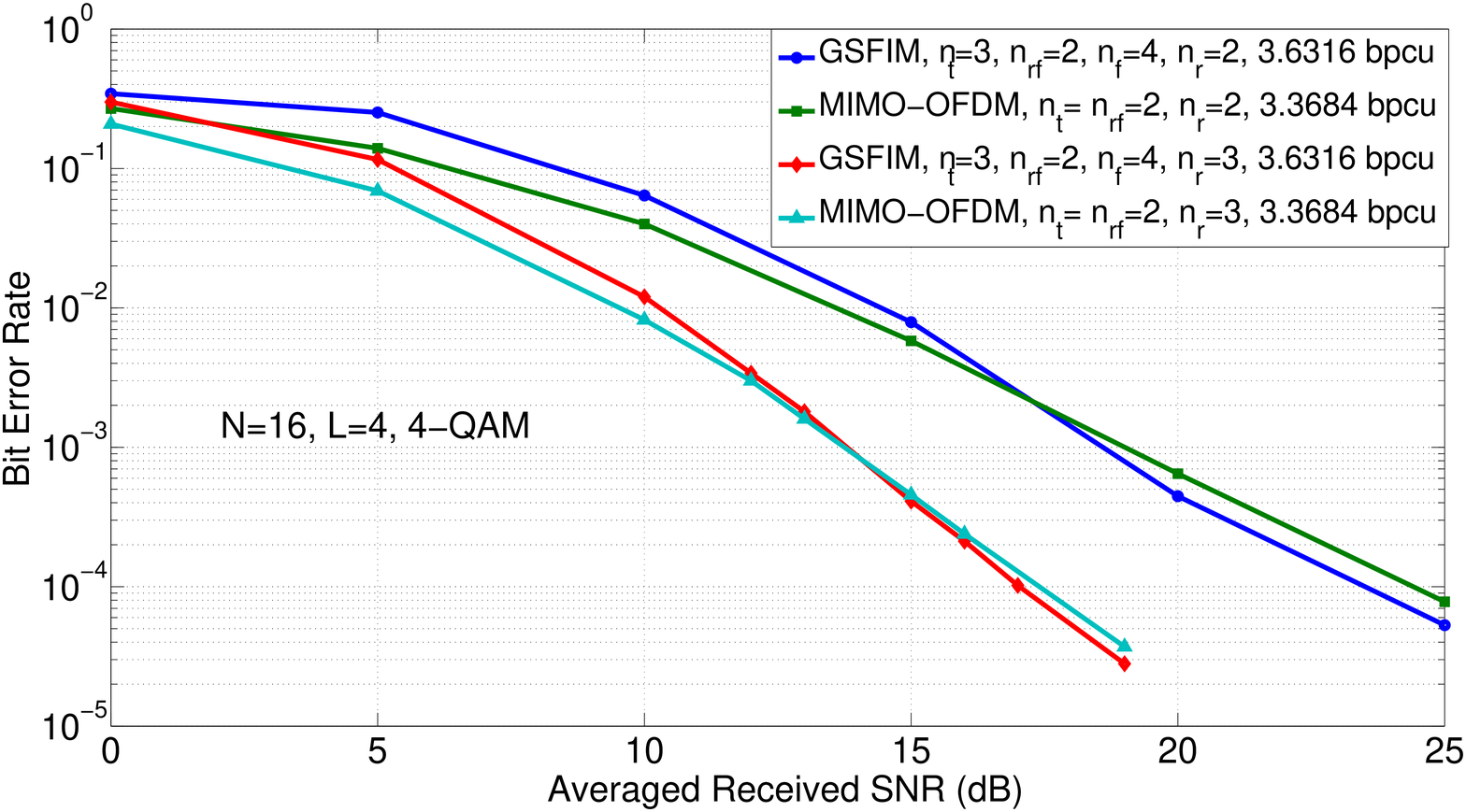}}
\caption{BER performance of GSFIM and MIMO-OFDM under ML detection.
(a) GSFIM with $n_t=3$, $n_{rf}=2$, $N=8$, $n_f=4$, $n_r=2,4$, $L=4$,
4-QAM, 3.1818 bpcu, and MIMO-OFDM with $n_t=n_{rf}=2$, $N=8$, $n_r=2,4$,
$L=4$, 4-QAM, 2.9091 bpcu.
(b) GSFIM with $n_t=3$, $n_{rf}=2$, $N=16$, $n_f=4$, $n_r=2,3$, $L=4$,
4-QAM, 3.6316 bpcu, and MIMO-OFDM with $n_t=n_{rf}=16$, $N=16$, $n_r=2,3$,
$L=4$, 4-QAM, 3.3684 bpcu.
}
\label{fig_gsm1b}
\vspace{-4mm}
\end{figure}

In Figs. \ref{fig_gsm1b}(a) and (b), we show the BER performance
of GSFIM in comparison with MIMO-OFDM under ML detection.
In Fig. \ref{fig_gsm1b}(a), the GSFIM system has $n_t=3$, $n_{rf}=2$,
$N=8$, $n_f=4$, $n_r=2,4$, 4-QAM, and the achieved rate is
$R_{\mbox{\scriptsize gsfim}}=3.1818$ bpcu.
The MIMO-OFDM has $n_t=n_{rf}=2$, $N=8$, $n_r=2,4$, 4-QAM, and the
achieved rate is $R_{\mbox{\scriptsize mimo-ofdm}}=2.9091$ bpcu.
In Fig. \ref{fig_gsm1b}(b), the GSFIM system has $n_t=3$, $n_{rf}=2$,
$N=16$, $n_f=4$, $n_r=2,3$, $L=4$, 4-QAM, and the achieved rate is
$R_{\mbox{\scriptsize gsfim}}=3.6316$ bpcu. The MIMO-OFDM system has
$n_t=n_{rf}=2$, $N=16$, $n_r=2,3$, $L=4$, 4-QAM, and the achieved
rate is $R_{\mbox{\scriptsize mimo-ofdm}}=3.3684$ bpcu. It is seen 
that in Figs. \ref{fig_gsm1b}(a) and (b), GSFIM has higher rates than 
MIMO-OFDM. In terms of error performance, while MIMO-OFDM performs 
better at low SNRs, GSFIM performs better at moderate to high SNRs. 
This performance cross-over can be explained in the same
way as explained in the case of GSIM in the previous section (Sec.
\ref{sec6_6}, Figs. \ref{fig6_9} and \ref{fig6_10}); i.e., at
moderate to high SNRs, errors in index bits are less likely and this
makes GSFIM perform better; at low SNRs, index bits and hence the
associated modulation bits are more likely to be in error making
MIMO-OFDM to perform better. Similar performance cross-overs
have been reported in the literature for single-antenna OFDM
with/without subcarrier indexing (e.g., \cite{scim2}), where it
has been shown that OFDM with subcarrier indexing outperforms
classical OFDM without subcarrier indexing at moderate to high
SNRs, whereas classical OFDM outperforms OFDM with subcarrier
indexing at low SNRs. The plots in Figs. \ref{fig_gsm1b}(a) and (b)
essentially capture a similar phenomenon when there are index bits
both frequency as well as spatial domains.

\section{Conclusions}
\label{sec_conc}
We introduced index modulation where information bits are encoded in the
indices of the active antennas (spatial domain) and subcarriers (frequency
domain), in addition to conveying information bits through conventional
modulation symbols. For generalized spatial index modulation (GSIM),
where bits are indexed only in the spatial domain, we derived the
expression for achievable rate as well as easy-to-compute upper and 
lower bounds on this rate. We showed that the achievable rate in GSIM
can be more than that in spatial multiplexing, and analytically
established the condition under which this can happen. We also proposed
a Gibbs sampling based detection algorithm for GSIM and showed that
GSIM can achieve better BER performance than spatial multiplexing. 
GSIM achieved this better performance using fewer transmit RF chains
compared to spatial multiplexing. 
For generalized space-frequency index modulation (GSFIM), where bits
are encoded in the indices of both active antennas as well as subcarriers,
we derived the achievable rate expression. Numerical results showed that
GSFIM can achieve higher rates compared to conventional MIMO-OFDM. Also,
BER results using ML detection showed the potential for GSFIM performing
better than MIMO-OFDM at moderate high SNRs. Low complexity detection 
methods for GSFIM can be taken up for future extension to this work.

\section*{Acknowledgment}
The authors would like to thank Mr. T. Lakshmi Narasimhan and Mr.
B. Chakrapani for their valuable contributions to the discussions 
on index modulation techniques. 

\bibliographystyle{IEEE}

\end{document}